\documentclass[glov3, smallextended, final]{svjour3}       
\pdfoutput=1
\smartqed  

\usepackage{mathptmx}      
\usepackage{lscape, graphicx, geometry, subfigure, amssymb, amsmath}
\usepackage[usenames,dvipsnames,svgnames,table]{xcolor}
\usepackage{setspace, textcomp, tabularx}
\usepackage{natbib}
\usepackage{hyperref}
\usepackage[nottoc]{tocbibind}
\hypersetup{breaklinks=true, colorlinks=true, citecolor=blue,linkcolor=black}
\journalname{The Astronomy and Astrophysics Review}
\begin{document}

\title{Radio observations of active galactic nuclei with mm-VLBI
}


\author{B.Boccardi \and
        T.P. Krichbaum \and 
        E. Ros \and
        J.A. Zensus 
}


\institute{B. Boccardi; T.P. Krichbaum; J.A. Zensus \at 
              Max-Planck-Institut f\"{u}r Radioastronomie, Auf dem H\"{u}gel 69, D-53121 Bonn, Germany  \\
              \email{bboccardi@mpifr.de}           
           \and
            E. Ros \at
            Max-Planck-Institut f\"{u}r Radioastronomie, Auf dem H\"{u}gel 69, D-53121 Bonn, Germany 
          \and
          Observatori Astron\`{o}mic, Universitat de Val\`{e}ncia, C.\ Catedr\'{a}tico Jos\'{e} Beltr\'{a}n 2, E-46980 Paterna, Val\`{e}ncia, Spain
          \and
          Departament d'Astronomia i Astrof\'{\i}sica, Universitat de Val\`{e}ncia, C.\ Dr.\ Moliner 50, E-46100 Burjassot, Val\`{e}ncia, Spain }

\date{Received: date / Accepted: date}

\maketitle

\begin{abstract}
Over the past few decades, our knowledge of jets produced by active galactic nuclei (AGN) has greatly progressed thanks to the development of very-long-baseline interferometry (VLBI). Nevertheless, the crucial mechanisms involved in the formation of the plasma flow, as well as those driving its exceptional radiative output up to TeV energies, remain to be clarified. Most likely, these physical processes take place at short separations from the supermassive black hole, on scales which are inaccessible to VLBI observations at centimeter wavelengths. Due to their high synchrotron opacity, the dense and highly magnetized regions in the vicinity of the central engine can only be penetrated when observing at shorter wavelengths, in the millimeter and sub-millimeter regimes. While this was recognized already in the early days of VLBI, it was not until the very recent years that sensitive VLBI imaging at high frequencies has become possible. Ongoing technical development and wide band observing now provide adequate imaging fidelity to carry out more detailed analyses.

In this article we overview some open questions concerning the physics of AGN jets, and we discuss the impact of mm-VLBI studies. Among the rich set of results produced so far in this frequency regime, we particularly focus on studies performed at 43\,GHz (7\,mm) and at 86 GHz (3 mm). Some of the first findings at 230 GHz (1 mm) obtained with the Event Horizon Telescope are also presented.

\end{abstract}
\newpage
\renewcommand\contentsname{}
\setcounter{tocdepth}{2}
\tableofcontents
  \large
\section{\large Introduction}
Astrophysical jets count among the most spectacular and powerful objects in the Universe. 
They are collimated outflows of plasma observed in a variety of astronomical sources -- young stellar objects (YSOs), 
X-ray binaries (XRBs), active galactic nuclei (AGN) and $\gamma$-ray bursts (GRBs) -- spanning many order of magnitudes both in the energy domain and in linear scale. Despite the diversity of environments where they can originate, all jets share some common features. Their power source is believed to be the gravitational potential of a compact and accreting central object \citep{1964ApJ...140..796S, 1969Natur.223..690L}, whose mass largely determines the scaling properties  \citep{1996Natur.382...47S,2003MNRAS.343L..59H,2003MNRAS.345.1057M,2004A&A...414..895F}. 
Moreover, all jets are magnetized, as they are, especially in the radio band, copious emitters of synchrotron radiation. Analyzing the similarities as well as the differences between the classes of jets is crucial for ultimately understanding the formation and propagation of the outflows, and the connection between the accretion properties and the jet activity \citep[e.g.,][]{2003NewAR..47..667M, 2010LNP...794.....B}.

AGN jets, powered by supermassive black holes at the center of some active galaxies, certainly form the most studied class. These highly collimated outflows, with opening angles of few degrees, propagate often undisturbed up to kiloparsec and sometimes megaparsec distances, and radiate over a broad interval of the electromagnetic spectrum. Most of their total power, varying in the range $10^{43}\,-10^{48}\,\rm{erg/s}$ \citep{2014Natur.515..376G}, is however not radiative, but carried in different forms. Close to the launching site it may be purely electromagnetic, while on larger scales it converts to mostly kinetic as the bulk flow accelerates \cite[e.g.,][]{2012bhae.book.....M}, reaching terminal Lorentz factors of the order of ten \citep{2016AJ....152...12L}. Ultimately, the energy gets dissipated in the form of radiation, giving rise to irregular structures of diffuse radio emission known as radio lobes, sometimes punctuated by compact hotspots. A beautiful representation of the large scale morphology 
in the radio galaxy Hercules A is shown in Figure 1. Although, according to the unification schemes \citep[e.g.,][]{1989ApJ...336..606B, 1995PASP..107..803U}, all jets are intrinsically elongated and two-sided as in Hercules\,A, many of the sources we can observe appear highly compact and asymmetric. This is due to strong relativistic and projection effects arising from the close alignment of the jet axis with our line of sight, which make the jet properties even more dramatic, but also more difficult to study.  
\begin{figure}[!h]
 \includegraphics[width=\textwidth]{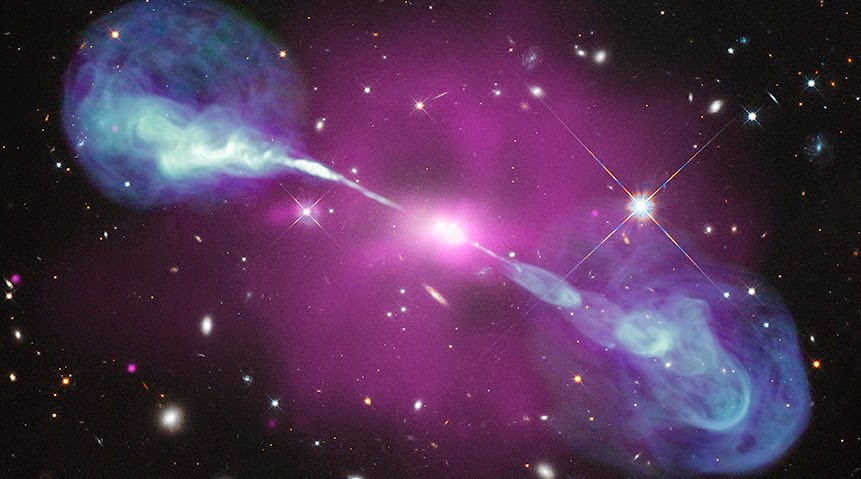}
 \caption{Composite image of the radio galaxy Hercules A. In blue is the radio emission associated with the jets and the lobes; in pink the X-ray emission from the heated surrounding gas; in white, orange, and blue the host galaxy and the background optical field. (Credit: X-ray: NASA/CXC/SAO, Optical: NASA/STScI, Radio: NSF/NRAO/VLA).}
\end{figure}
\subsection{\large The quest for angular resolution}
Starting in 1963, when Maarten Schmidt revealed the extragalactic nature of the radio source 3C\,273 \citep{1963Natur.197.1040S}, astronomers have extensively investigated the physical processes which could lead to those tremendous energy outputs. The puzzle became even more complicated when the emission was observed to vary on extremely short timescales $t_{\rm var}$ \citep[see][and references therein]{1970Natur.227.1303R}. Based on the light travel time argument, the sizes of the emitting regions $l\leq c\cdot t_{\rm var}$ (with $c$ indicating the speed of light) were determined to be as small as few light months. The extreme compactness implied was not only a challenge for theorists, but also for observers. Expressed in angular dimensions, the relevant scales for an AGN jet are in fact of the order of the milli-arcsecond or smaller (for typical distances up to few Giga-parsecs), reaching far beyond the diffraction limit of a single telescope. 

Already in the 1940s, radio astronomers had been looking for smart solutions aimed at increasing the resolving power of their instruments. 
This goal was brilliantly achieved with the development of aperture synthesis \citep{1946Natur.158..339R, 1960MNRAS.120..220R}, acknowledged by the first Nobel prize for astronomical research, in 1974. Aperture synthesis is an interferometric technique aimed at synthesizing a very large effective aperture from an array of telescopes. This elegant method is applied in its utmost complexity in very-long-baseline interferometry (VLBI), in which the astronomical signal collected by physically unconnected telescopes can be combined. The angular resolution of an interferometer is proportional to the ratio $\lambda/b_{\rm max}$, where $\lambda$ is the wavelength of the radiation and $b_{\rm max}$ is the maximum baseline length, i.e., the maximum (projected) separation between two elements of the array. The major VLBI arrays currently in use, the Very Long Baseline Array\footnote{After recent structural changes within the National Radio Astronomy Observatory (NRAO), the VLBA is now 
an independent facility operated by the 
Long Baseline Observatory (LBO - https://www.lbo.us)} (VLBA) and the European VLBI Network (EVN), are characterized by maximum baselines of the order of 10000 kilometers, and perform most of their observations at 
centimeter 
wavelengths. Their operation is crucial in the investigation of compact radio sources, and has advanced significantly our knowledge of the physical conditions of the plasma flow on milli-arcsecond scales, i.e., on projected spatial extents of the order of few parsecs.

However, the quest for a deeper understanding of these objects is not over. Based on observational and theoretical grounds, the fundamental mechanisms involved in the energy production and in the very formation of the jet outflow are expected to take place on even smaller scales, comparable to the Schwarzschild radius $R_{\rm S}=2GM_{\rm BH}/c^2$ ($G$ is the gravitational constant and $M_{\rm BH}$ is the black hole mass) of the supermassive black hole. Radio interferometric observations capable of probing such scales are therefore decisive for obtaining a complete picture of the AGN phenomenon, particularly if used in synergy with high-frequency studies, in the optical, X-ray, and $\gamma$-ray bands \citep[e.g.][]{2005MmSAI..76..168M}.  

When aiming at improving the angular resolution of a radio interferometer, two main approaches can be followed. The first consists in increasing the maximum baseline by having one or more telescopes orbiting in space. After some pioneer experiments performed in the 80's, space-VLBI was successfully realized with the VLBI Space Observatory Program (VSOP) \citep{2000PASJ...52..955H} and, more recently, with the RadioAstron mission \citep{2013ARep...57..153K}. In the second approach, ground telescopes can be equipped with receivers operating at shorter wavelengths, in the millimeter or sub-millimeter bands. Both methods enable resolutions of a few tens of micro-arcseconds, which in the closest objects translate into linear sizes as small as few Schwarzschild radii. However, since the nuclear environment is dense and highly magnetized, a truly sharp view of radio cores in AGN can only be obtained by penetrating the opacity barrier shrouding them. Both synchrotron and free-free opacity can affect 
significantly the cm-wave emission, but are expected to be much reduced in the millimeter band. Millimeter VLBI combines in a unique manner a high spatial resolution with a spectral domain where source-intrinsic absorption effects vanish, and is therefore ideally suited for the imaging of the still unexplored regions in the vicinity of the black hole.  

In this article, we discuss some open questions concerning the physics of compact radio sources, and the impact of mm-VLBI observations towards a more detailed physical understanding. In Section 2 we briefly summarize the historical development of this sophisticated technique and we report on the capabilities of current mm-VLBI arrays. Section 3 is intended to provide the reader with an overview of the basic properties of jets as inferred from VLBI studies, and of the theoretical models that best describe them. In Section 4 we expand the discussion on the scientific topics that mm-VLBI can address and we present the main results produced so far. We conclude with an outlook of the future developments and goals of the mm-VLBI science (Sect. 5).

\section{\large mm-VLBI arrays}
\subsection{\large General concepts}

Very-long-baseline interferometry (VLBI) is an elegant observing technique which, as of this writing, provides the highest possible angular resolution in astronomy \citep[see][for the highest resolution image produced to date]{2016ApJ...817...96G}. In the following part of this section, we will detail the capabilities of VLBI arrays operating in the millimeter regime. First, however, we wish to introduce some necessary terminology and fundamental parameters which define the performance of radio telescopes and interferometers. The following description is by no means complete, and the reader is referred to the specialized textbooks on interferometry and synthesis imaging \citep[e.g.,][]{2017isra.book.....T}.

A radio interferometer is an instrument which enables to combine the radio waves coming from an astronomical object to form interference fringes. By correlating the signals collected simultaneously by each telescope forming the array, radio interferometers measure the \textit{complex visibility function} $V(u,v)$, which is the (noise-corrupted) Fourier transform of the brightness distribution of the sky. The $(u,v)$ coordinates define, in units of wavelength, the East-West and the North-South component of each baseline projected in the sky, as seen from the source. Thus, the $(u,v)$-plane contains information about the existence or absence of a visibility measurement in a certain point. The filling of the $(u,v)$-plane, i.e., the $(u,v)$-\textit{coverage}, is by definition incomplete for any interferometer, and can be improved by adding more telescopes to the array and by increasing the on-source time up to 12 hours, so that the Earth rotation enables a single baseline to sample a full track (an ellipse) in 
the $(u,v)$-plane. Each baseline of projected length $b$ is characterized by its own fringe pattern, and is only sensitive to source structures 
on scales comparable to the fringe spacing $\lambda/b$. Therefore, the better the sampling of the $(u,v)$-plane, the more reliable will be the reconstruction of the sky brightness distribution. The smallest angular scale an interferometer can probe, its resolution, coincides, in general, with the diffraction limit $\sim\lambda/b_{\rm max}$, where $b_{\rm max}$ is the maximum baseline length. 

While the angular resolution achieved at a given wavelength depends, in principle, only on the geometrical properties of the array, the sensitivity of interferometric observations is largely dependent on the characteristics of the single telescopes forming the array and of the recording systems in use. 
Given an array formed by $N$ telescopes, the sensitivity of a single telescope with geometrical area $A$ is quantified by the \textit{system equivalent flux density} $\rm SEFD$, a parameter which accounts for the combined sensitivity of the antenna and of the receiving system. The $\rm SEFD$\,$\rm [Jy]$ is given by the ratio $T_{\rm sys}/g$, where the \textit{system temperature} $T_{\rm sys}$\,$\rm [K]$ is the sum of all the noise contributions, both from the electronics and from the astronomical source, while the \textit{gain} $g$\,$[\rm K/Jy]$ quantifies the radio noise power received from a source of unit flux density. The gain is a measure of the antenna efficiency $\eta$, an elevation-dependent parameter which determines the effective collective area $A_{\rm eff}=\eta A$ of the dish. At short radio wavelengths, the efficiency is especially limited by the antenna surface accuracy. For instance, a surface accuracy $s=\lambda/16$ implies a reduction of the effective collecting area by a factor $\eta_s=\exp(
-4\pi s/\lambda)^2\simeq0.5$.

The astronomical signal is extremely weak with respect to the noise produced by the receiving system, and the signal-to-noise ratio must then be increased by averaging over a large number of samples $n_{\rm s}$. However, the number of samples cannot be arbitrarily high, being limited by the bandwidth of the signal $\Delta\nu$. According to the Nyquist sampling theorem, samples taken over time intervals $\Delta t$ shorter than $1/2\Delta\nu$ are not independent. Therefore the number of samples $n_{\rm s}=2\Delta\nu t$ has to be increased by increasing the observing time $t$ and/or by extending the observing bandwidth $\Delta\nu$. According the central limit theorem, averaging over a large number of samples will reduce the rms noise $\sigma$ by a factor of $\sqrt{n_{\rm s}}$, which allows us to write the \textit{radiometer equation} as:
 \begin{equation}
 \sigma=\frac{\sqrt{2}T_{\rm sys}}{\sqrt{2\Delta\nu t}}=\frac{T_{\rm sys}}{\sqrt{\Delta\nu t}}.
 \end{equation}

Among the possible ways to achieve a higher sensitivity, increasing the bandwidth is the most cost-effective. For this reason, the technical improvements in VLBI, and in radio interferometry in general, have been especially oriented towards the development of wide bandwidth receivers and recording systems.

In VLBI mode, radio signals are digitized and recorded at each telescope on magnetic tapes, together with extremely precise time stamps usually provided by an hydrogen maser clock. The digitization process has also an impact on sensitivity, and reduces the signal to noise ratio by a factor $\eta_{\rm c}\leq1$, called VLBI system efficiency, with respect to an ideal analog recording. The VLBI system efficiency depends on the number of bits used to represent each sample, with $\eta_{\rm c}\sim0.63$ for a 1-bit representation and $\eta_{\rm c}\sim0.87$ for a 2-bit representation.The sensitivity provided by a certain digital recording system is therefore determined by the data rate, expressed in $\rm{bit/s}$.

Ultimately, the theoretical thermal noise $\sigma_{\rm{im}}$ of a VLBI image depends both on the SEFD of each telescope and on the data rate, and can be expressed in units of $\rm{Jy/beam}$ as

\begin{equation}
\sigma_{im}=\frac{1}{\eta_{\rm c}}\frac{SEFD^*}{\sqrt{N(N-1)\Delta\nu t_{\rm int}}},
\end{equation}
where $SEFD^*=\sqrt{\sum_{i,j=1}^{N;i<j}(SEFD^i\times SEFD^j)}$ and $t_{\rm int}$ is the total integration time on source in seconds.

\subsection{\large A brief history and current status}

The possibility of conducting VLBI experiments at millimeter wavelengths was first explored in the late 70's. 
Only a few years later, in the early 80's, pioneering experiments were successfully performed through the observation of the strong radio source 3C\,84 at 86\,GHz (3\,mm) \citep{1983Natur.303..504R} and 43\,GHz (7\,mm) \citep{1985ESOC...22..157M}. The choice of these observing bands was naturally determined by the behavior of atmospheric transmission in the radio window. As shown in Fig. 2, strong opacity spikes due to molecular absorption in the troposphere (mainly from water vapor and oxygen) overlay the general trend of steeply increasing opacity as a function of frequency, and bracket three main observing windows at around $35$ $\rm GHz$, $90$ $\rm GHz$, and $135$ $\rm GHz$. 
\begin{figure}
\centering
 \includegraphics[width=0.8\textwidth]{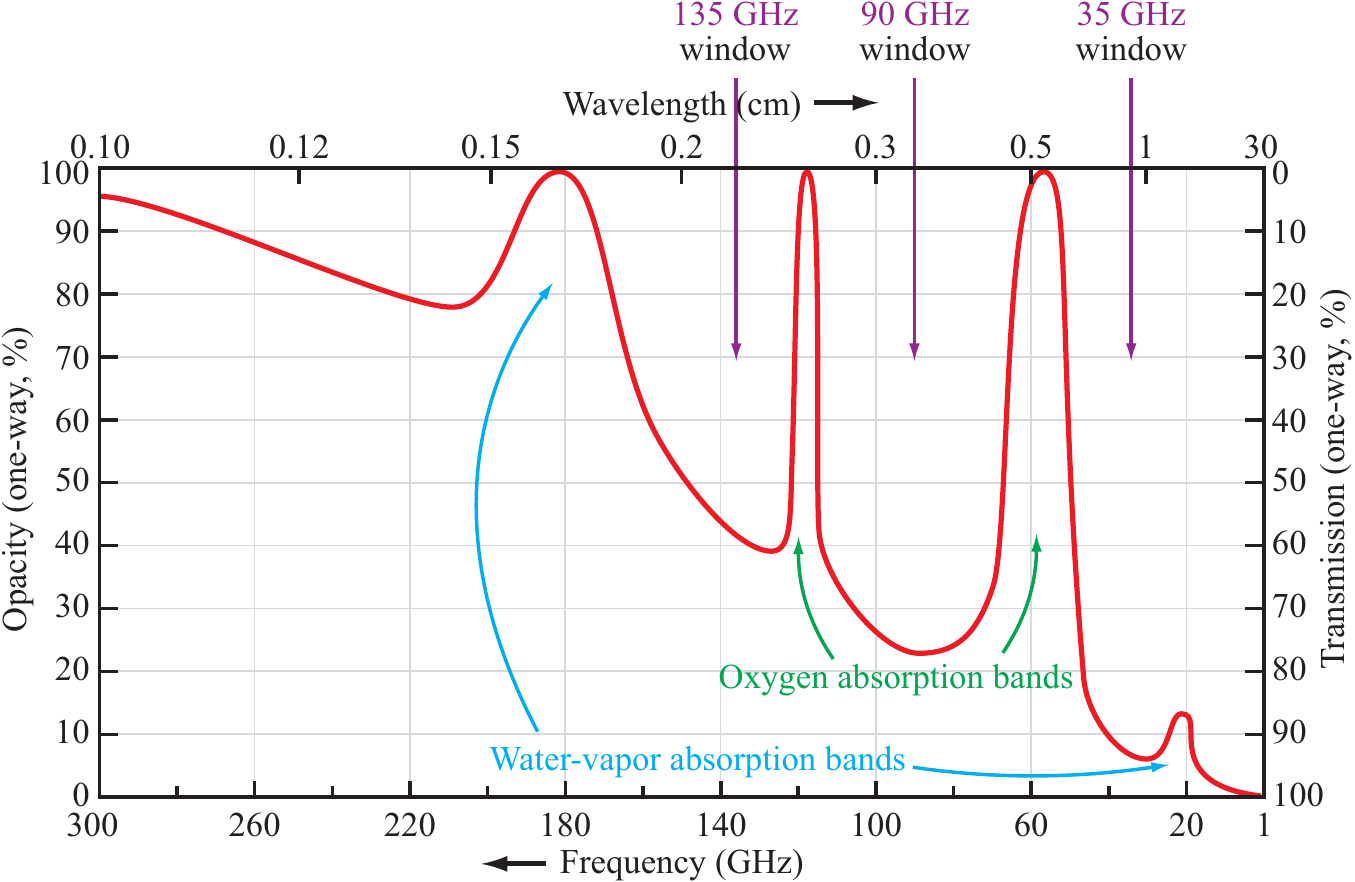}
 \caption{Percentage transmission through the Earth\textquoteright s atmosphere, in the vertical direction, under clear sky conditions \citep{ulaby2014microwave}. Strong molecular absorption is caused by the water vapor and the oxygen in the troposphere.}
\end{figure}

Since those first detections, the performance of mm-VLBI arrays has been steadily improving, until a sensitive imaging \citep[with dynamic range $>1000$, e.g.,][]{2016A&A...585A..33B} has become possible. The road through the technical development has been challenging, and sometimes rough. The main limitation in interferometric experiments at millimeter wavelengths is the reduced phase stability, resulting in coherence losses. In VLBI experiments, phase fluctuations can result from the contribution of several factors, e.g., inaccuracies in the hydrogen maser frequency standards or in the local oscillator systems of the receiver, antenna-pointing errors, uncertainties in the antennas positions \citep[see e.g.,][]{1981ITIM...30..283R}. At high radio frequencies, however, phase drifts are mostly of atmospheric origin \citep[see][for a discussion]{1984RaSc...19.1552R}, being primarily caused by water vapor in the troposphere. High altitude and/or dry sites are the optimal choice for minimizing the degradation of 
the signal in 
such observations. While excellent instruments, like the Plateau de Bure interferometer or the Atacama Large Millimeter Array (ALMA), were built ¨ad hoc¨ in optimal sites in the following years, initial millimeter arrays could only count on the existing telescopes, most of which were conceived for operating at centimeter-waves. This, combined with the poor $(u,v)$-coverage, limited bandwidths and inaccurate calibration procedures, has severely hindered the imaging capabilities of past mm-VLBI arrays. 

An important step forward was made with the foundation of global arrays operating at 86\,GHz, the Coordinated Millimeter VLBI Array (CMVA) \citep{1995AAS...187.1212R}, established in 1995, and its successor, the Global Millimeter VLBI Array (GMVA)\footnote{http://www3.mpifr-bonn.mpg.de/div/vlbi/globalmm/} \citep{2006evn..confE...2K}, whose activity is ongoing since 2003. The participation of a larger number of telescopes ($>10$), the inclusion of sensitive antennas like the IRAM telescopes (Pico Veleta and the phased Plateau de Bure) or the Effelsberg 100-m, and the remarkable increase in bandwidth (by a factor $>$10) gradually led to an improvement of the performance. This fact can be immediately recognized when considering the detection rate of 86\,GHz VLBI surveys conducted between 1997 and 2008. This increased from a percentage of less than $25\%$ obtained in the first campaigns \citep{1997mvlb.work...53B, 1998AJ....116....8L, 1998A&AS..131..451R} to more than $90\%$ achieved in the latest \citep{2000A&A.
..364..391L, 2008AJ....136..159L, 2016cosp...41E.710G}.

\begin{table}
\centering
\caption{Properties of the antennas operating at 86\,GHz. Col. 1: Name of the telescope. Col. 2: Its location. Col. 3: Effective diameter in meters. Col. 4: Typical value of the system equivalent flux density (SEFD), which is a measurement of the telescope sensitivity. Lower values indicate a better performance.}
\label{my-label}
\begin{tabular}{llll}
\hline\noalign{\smallskip}
Station    & Location & Effective diameter $\rm [m]$ & SEFD $\rm [Jy]$ \\ 
\noalign{\smallskip}\hline\noalign{\smallskip}
Effelsberg          & Germany                     & 80              & 1000  \\ 
Onsala              & Sweden                      & 20              & 5100  \\
Plateau de Bure	    & France                      & 34              & 820   \\ 
Pico Veleta         & Spain                       & 30              & 650   \\ 
Yebes               & Spain                       & 40              & 1700  \\ 
Mets\"{a}hovi       & Finland                     & 14              & 17000 \\
Green Bank          & United States               & 100             & 140   \\ 
VLBA ($\times8$)    & United States               & 25              & 2500  \\
KVN ($\times3$)     & South Korea                 & 21              & 3200  \\
ALMA                & Chile                       & 85              & 60  \\
\noalign{\smallskip}\hline
\end{tabular}
\end{table}

\begin{figure}[!ht]
\centering
 \includegraphics[width=0.95\textwidth]{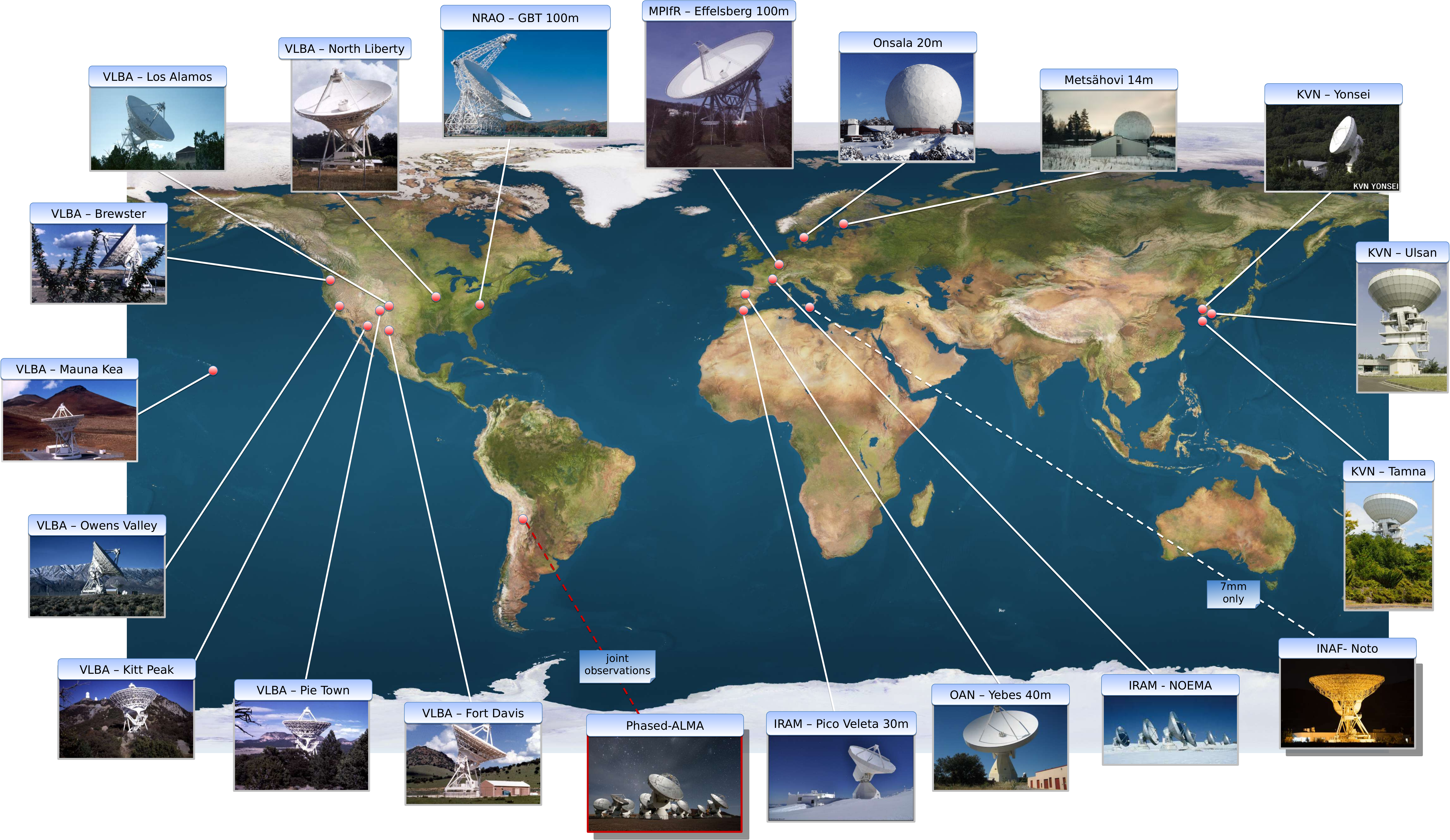}
 \caption{The telescopes forming the Global Millimeter VLBI Array (GMVA), operating at the frequency of 86\,GHz. Credit: Helge Rottmann.}
\end{figure}

Currently, the GMVA comprises up to 18 telescopes (Effelsberg, Onsala, Plateau de Bure, Pico Veleta, Yebes, Mets\"{a}hovi, Green Bank, eight VLBA stations and the three telescopes forming the Korean VLBI Network - KVN) spread over three continents (Europe, America, and Asia -- Table 1, Fig. 3). The phased ALMA, located in Chile, has also participated for the first time in mm-VLBI experiments in April 2017. To date, most of the GMVA stations are observing at a bandwidth of 512 MHz (which corresponds to a data rate of 2048 Mbit/s), and further upgrades are expected in the near future thanks to the fast development of digital VLBI recording systems.  Global VLBI observations at 86\,GHz can achieve a typical angular resolution of 50-70 micro-arcseconds, and an array sensitivity of $\sim$0.9 milli-Jansky/hour.  

At frequencies higher than 86\,GHz, atmospheric effects become even more dominant, and VLBI experiments are plagued with additional difficulties. Nevertheless, fundamental science goals (Sect. 4.5) have driven the technical development also in this frequency regime. A first experiment at 223\,GHz (1.4 mm) was successfully performed already in 1989 \citep{1990ApJ...360L..11P}, with a single-baseline fringe detection of 3C\,273. Subsequent observations involving telescopes like Pico Veleta, Plateau de Bure, CARMA (Combined Array for Research in Millimeter-Wave Astronomy), the SMTO (Sub-millimeter telescope) and, since 2012, also APEX (Atacama Pathfinder Experiment) confirmed the feasibility of 1mm-VLBI \citep[e.g.,][]{1997A&A...323L..17K, 1998A&A...335L.106K, 2008Natur.455...78D, 2015A&A...581A..32W}, achieving the detection of several other sources also on transatlantic baselines. Much of the development of 1mm-VLBI is driven at present by the Event Horizon Telescope (EHT)\footnote{http://www.
eventhorizontelescope.org/} project. 
When completed, the EHT will be able resolve the black hole surroundings on scales comparable with the event horizon. The expected resolution is of 20--30 $\mu$as at 230 GHz and of 13--20 $\mu$as at 345 GHz. Due to the large apparent size of their event horizon, Sagittarius A* and M\,87 are the best candidates for achieving such a goal, and therefore are considered as primary targets for EHT observations.  

\section{\large Extragalactic jets as seen by VLBI}

In VLBI images, the radio emission from an AGN can be typically ascribed to a compact, bright, and unresolved feature called the ``core'', and to a one-sided jet emanating from it (Fig. 4). This classical morphology is the result of selection effects, involving both the relativistic nature of the flow and the sensitivity of VLBI arrays. 

Similarly to Hercules\,A (Fig. 1), all jets are thought to be characterized by a roughly symmetric two-sided structure. However, due to the relativistic nature of the flow, the emission from the side approaching the observer can be highly enhanced as a consequence of relativistic Doppler boosting. By applying the Lorentz transformations, it can be shown that the observed flux density $S_{\rm o}$ differs from the intrinsic one $S_{\mathrm{e}}$ by a factor $\delta^{n-\alpha}$, where $\delta$ is the Doppler factor, $\alpha$ is the spectral index, and $n$ is a parameter varying between 2 and 3 \citep[see][]{1979Natur.277..182S}. The Doppler factor depends on the speed of the flow $\beta=v/c$, which determines the bulk Lorentz factor $\Gamma=1/\sqrt{(1-\beta^2)}$,  and on the jet orientation, defined by the angle $\theta$ between the direction of propagation of the outflow and the line of sight of the observer:
\begin{equation}
\delta=\frac{1}{\Gamma(1-\beta\cos\theta)}.
\end{equation}
Given a population of jets oriented randomly and spanning a certain range of Lorentz factors, it is clear from Eq. 3 that Doppler boosting increases the chances to detect the fastest objects and/or those oriented at small angles \citep[naturally, there is also a dependence on the intrinsic luminosity function of the jet population, see e.g.][]{1997ApJ...476..572L}. The frequent observation of superluminal motion, i.e., of apparently faster-than-light speeds $\beta_{\rm app}$ of the plasma features in the jet, confirms the selection bias affecting the jets samples. This purely geometrical phenomenon, predicted by Martin Rees in 1966 \citep{1966Natur.211..468R} and observed for the first time in 3C\,273 \citep{1969Natur.224.1094G}, occurs more prominently in fast flows seen at small angles, as described by the relation 
\begin{equation}
\beta_{\rm app}=\frac{\beta\sin\theta}{1-\beta\cos\theta}. 
\end{equation}
\begin{figure}
\centering
\includegraphics[width=0.75\textwidth]{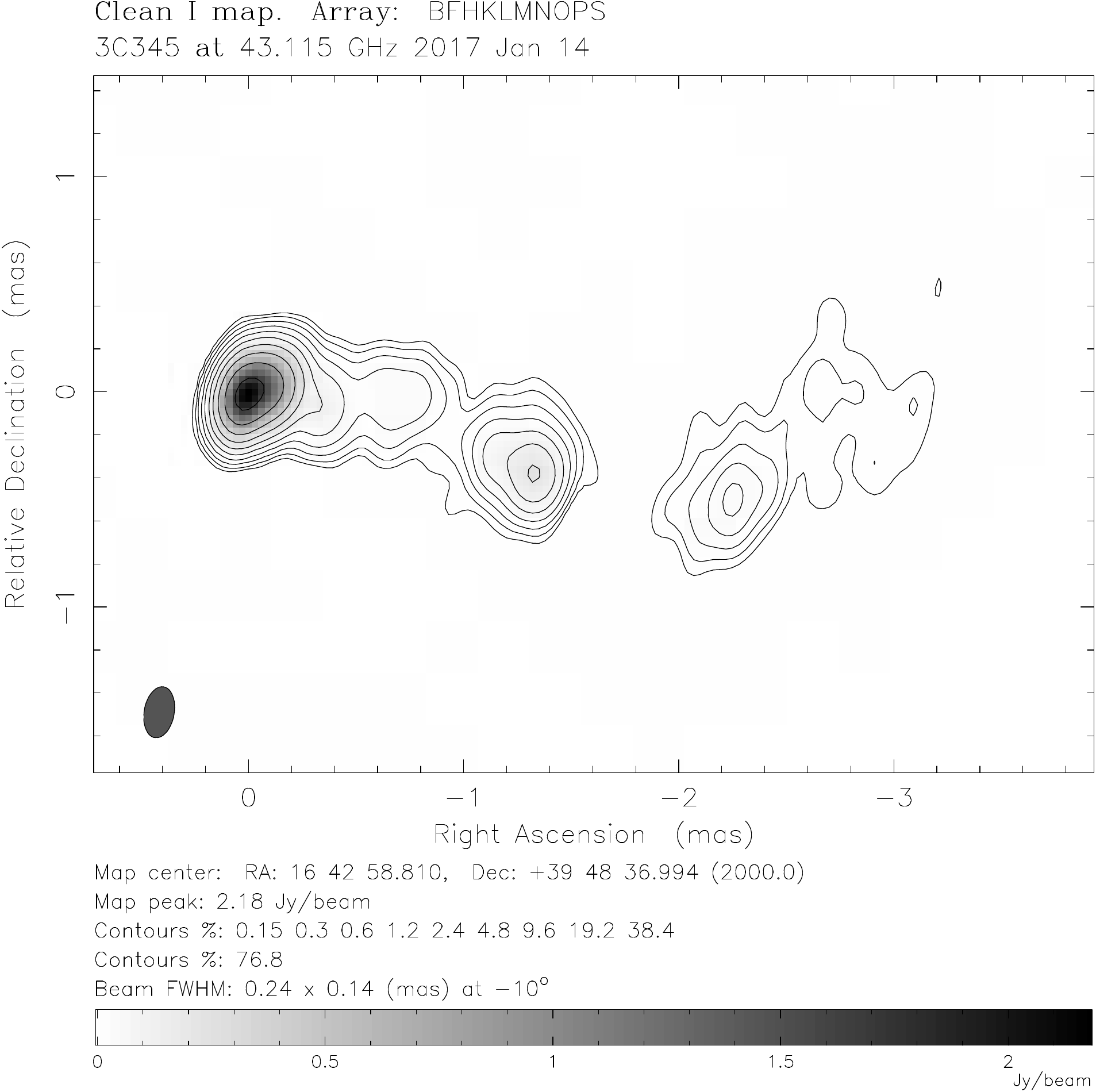}
 \caption{A 43\,GHz image of the blazar 3C\,345 from VLBA observations in January 2017. The source shows a typical core-jet structure, with bright knots punctuating the jet emission. The image was produced by the authors from the data available at the public archive of the VLBA-BU Blazar Monitoring Program (VLBA-BU-BLAZAR; http://www.bu.edu/blazars/VLBAproject.html).}
\end{figure}

Another selection effect in VLBI is related to sensitivity. It is not sufficient for an astronomical source to be a strong radio emitter to be seen by VLBI. Only ``high intensity'' objects, i.e., with high brightness temperature $T_{\rm b}$, are suitable targets.  The minimum brightness temperature $T^{\rm min}_{\rm b}$ detectable by an interferometer depends on the flux density $S_{\nu}$, the baseline length $b$, and the Boltzmann constant $k$ as
\begin{equation}
T^{\rm min}_{\rm b}=\frac{2}{\pi k} b^2 S_{\rm \nu},
\end{equation}
and is typically in the range $10^6-10^8$ $\rm K$. The latter implies that both thermal emission ($T_{\rm B}<10^5$ $\rm K$) and non-thermal emission of insufficient compactness (e.g., from the radio lobes) are completely resolved out in a VLBI observation. In conclusion, the samples targeted by VLBI are dominated by the highly boosted compact objects, also known as \textit{blazars}\footnote{According to the unification schemes \citep{1995PASP..107..803U}, blazars come in two flavors: FSRQs (Flat Spectrum Radio Quasars), which show broad emission lines and powerful, highly collimated jets, and BL Lacs, which lack emission or absorption features and have less powerful jets. The parent population of jets oriented at larger viewing angles is formed by the radio galaxies, Fanaroff-Riley II and Fanaroff-Riley I respectively \citep{1974MNRAS.167P..31F}.}. While studies of blazars may provide an incomplete view of the general jet phenomenon in AGN, these sources represent a unique cosmic laboratory for 
investigating the physics of relativistic plasmas in extreme conditions. 
\subsection{\large Where do we stand?}
The physical conditions of the plasma forming a jet evolve significantly as the jet propagates from the central engine through the medium. According to the current paradigm, it is possible to identify four distinct regions (Fig. 5): launching, acceleration and collimation, kinetic-flux dominated, and dissipation. 

Jets are thought to be launched in the immediate vicinity of the supermassive black hole, at distances of $\leq 10^2$\,$R_{\rm S}$ \citep{2001Sci...291...84M}. In the launching region, the plasma is channeled through the action of strong magnetic fields, which extract part of the energy stored in the accretion disk \citep{1982MNRAS.199..883B} and/or in the rotating black hole \citep{1977MNRAS.179..433B}. The composition of the loaded matter is not well determined: the jet may consist of a normal electron-proton plasma \citep[e.g.,][]{1993MNRAS.264..228C} or it may be formed by light particles only, electrons and positrons \citep[e.g.,][]{1996MNRAS.283..873R}. 

This purely electromagnetic flow is then accelerated and collimated. In the second region, reaching up to scales of $10^3-10^5$ $R_{\rm S}$ ($\sim$sub-parsec to parsec) \citep{2003ApJ...596.1080V, 2004ApJ...605..656V}, the initially broad stream is rapidly focused thanks to the 
confinement provided by the magnetic field and/or by the external medium, and is accelerated to relativistic speeds by magnetic pressure gradients \citep[e.g.,][]{1997MNRAS.288..333S,2007MNRAS.380...51K, 2009ApJ...698.1570L}. At the end of this process, a large part of the magnetic energy has been converted to kinetic energy. The kinetic-flux dominated jet extends between $10^5-10^9$ $R_{\rm S}$ (parsecs to kiloparsecs). In this region, the magnetic field is expected to become dynamically unimportant, and the jet can be appropriately described by the physical laws of gas dynamics \citep{1988ApJ...334..539D}. Being subject to hydrodynamic shocks \citep{1979ApJ...232...34B, 1985ApJ...298..114M} and plasma instability \citep{1972MNRAS.156...67B, 1979ApJ...234...47H}, the flow ultimately looses its collimation and dissipates its energy in the form of radiation. Within the lobes formed at distances $\geq 10^9$\,$R_{\rm S}$ ($>$kiloparsecs), compact hotspots are often observed, suggesting that part of the plasma can still be highly relativistic on large scales \citep{2004ApJ...604L..81G}. 

\begin{figure}[!ht]
\centering
 \includegraphics[width=0.95\textwidth]{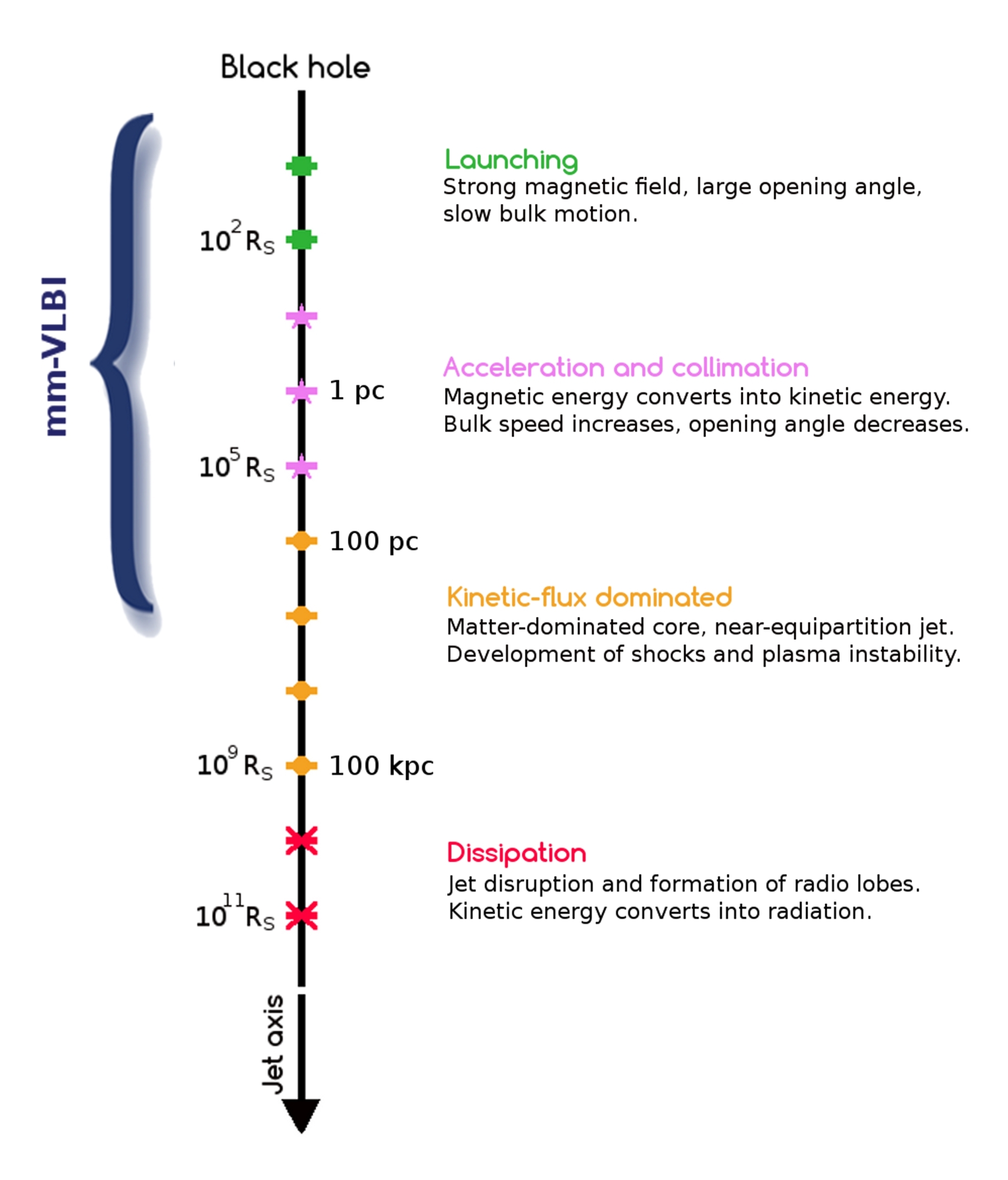}
 \caption{Schematic view of the main regions of a relativistic jet, according to the current paradigm for magnetically-driven flows. The radial separation from the black hole is represented in a logarithmic scale, in units of Schwarzschild radii $R_{\rm S}$. We also report the corresponding distance in units of parsecs ($\rm{pc}$) for a black hole mass of $10^9$\,$M_{\odot}$. The extension of each region is approximate, and may vary in different jets. VLBI observations at millimeter wavelengths are suited for probing the magnetically dominated jet base and the transition to the kinetic-flux dominated region. }
\end{figure}

Over the past 40 years, radio interferometric observations have provided us with a detailed description of the two outermost regions. In particular, VLBI studies at centimeter wavelengths have significantly improved our understanding of parsec-scale flows (also known as compact jets). Both statistical studies of large samples -- first among all, those performed at 2cm within the \textit{MOJAVE} program\footnote{http://www.physics.purdue.edu/astro/MOJAVE/} \citep{2009AJ....137.3718L}  -- and detailed analyses of single targets were able to address several key topics. How fast does the plasma flow? What is the energy balance between the magnetic field and the particles? Is the magnetic field ordered? What are jets made of? And what is that we really observe with VLBI? Providing a complete summary of these results is a challenging task, and beyond the scope of this review. The interested reader is referred to the review articles from \cite{1997ARA&A..35..607Z} and \cite{2006AIPC..856....1M}. In the following 
we discuss some highlights, with the aim of better contextualizing the scientific questions that are more relevant for  mm-VLBI studies.

Centimeter VLBI imaging has shown that a large fraction of the parsec-scale emission is produced by the core component, the upstream feature from which the jet appears to emanate. The radio core, characterized by a flat spectrum and weak polarization and usually located at parsec-distances from the black hole, is classically interpreted as marking the transition between synchrotron self-absorbed regions and optically thin regions \citep{1979ApJ...232...34B, 1981ApJ...243..700K}. The flatness of the spectrum is explained as the result of the superposition of different synchrotron self-absorbed components in a conical geometry \citep[see also][]{1996ASPC..100...45M}. In this scenario, the synchrotron opacity associated with the core component implies that its position is not fixed, but has a frequency-dependence. Specifically, the emission should be detected at progressively smaller distances from the central engine with increasing observing frequency. This effect, known 
as \textit{core-shift}, is observed in many jets \citep{1984ApJ...276...56M,1998A&A...330...79L}, although not in all \citep{2012A&A...545A.113P}. 

The emission from the jet is largely ascribed to the development of relativistic hydrodynamic shocks and plasma instability.
The Kelvin-Helmoltz instability \citep{1976MNRAS.176..443B, 1976MNRAS.176..421T, 1978A&A....64...43F}, formed in the presence of velocities shears between two fluids, is likely to dominate the emission and shape the flow on hectoparsec and larger scales \citep{2000ApJ...533..176H, 2001Sci...294..128L}.
Shocks, instead, are a distinctive feature of parsec-scale flows \citep{1979ApJ...232...34B, 1985ApJ...298..114M}. Indeed, most of the jets imaged with VLBI do not appear as continuous flows, but can be well modeled as a sum of discrete features, known as \textit{blobs} or \textit{knots}. 
Shocks originate either from pressure mismatches at the boundaries with the external medium or from major changes at the base of the flow (e.g., new plasma ejections or directional changes). They are sites of efficient particle acceleration through the Fermi mechanism \citep[e.g.,][]{1978MNRAS.182..147B} and of local amplification of the magnetic field. Accordingly, strong variability of the emission and enhanced polarization can often be associated with shocked regions in VLBI images of jets. Shocks are observed to move superluminally, with apparent speeds as high as ${\sim}50$\,$c$, but more commonly below $10$\,$c$ \citep{2016AJ....152...12L}. 

In addition to moving features, jets are also punctuated by bright stationary spots \citep[e.g.][]{2004ApJ...609..539K,2009AJ....137.3718L, 2012ApJ...752...92A}.
These have generally been interpreted as standing shocks, formed, e.g., as a consequence of prominent recollimation events \citep[e.g.,][]{1988ApJ...334..539D, 1995ApJ...449L..19G}. Stationary features in the vicinity of the jet base are thought to play an important role for the production of non-thermal high-energy emission in AGN \citep[e.g.,][]{2010ApJ...715..355L, 2010MNRAS.401.1231A}, and may mark fundamental transition regions of the jet. For instance, the presence of a recollimation shock may be expected at the end of the acceleration and collimation region, when the jet becomes causally disconnected from the central engine \citep[e.g.,][]{2010ApJ...723.1343P, 2014ApJ...787..151C}. 

In some objects, particularly in those not showing an appreciable core-shift, the VLBI core itself may coincide with such a recollimation feature or, more generally, with the first and brightest standing shock developing in the flow \citep{2008ASPC..386..437M, 2009ASPC..402..194M}. The nature of the radio core, still debated, is clearly a crucial element to understand, and much of the recent theoretical work has been aimed at identifying the dissipative events which could account for the broadband emission, variability and polarization properties observed in its vicinity \citep[e.g.,][and references therein]{2009ApJ...704...38S, 2015MNRAS.450..183S}. Radio observations indicate that, whichever mechanism is taking place, it must lead to very efficient energy dissipation. This is inferred, for instance, from measurements of the brightness temperature of individual components in VLBI images, obtained after estimating their flux density $S_\nu$ and angular dimension $d$ as:
\begin{equation}
 T_{\rm b}=1.22\cdot10^{12}\frac{S_\nu(1+z)}{d^2\nu^2}.
\end{equation}

At centimeter wavelengths, the intrinsic brightness temperature $T_{\rm 0}=T_{\rm b}/\delta$ of the core reaches \citep{1998AJ....115.1295K, 2006ApJ...642L.115H} and sometimes exceeds \citep{2016ApJ...820L...9K} the maximum theoretical value expected 
for a synchrotron self-absorbed component. This limit is posed by the onset of the \textit{inverse Compton catastrophe}, i.e., by the strong enhancement of the inverse Compton cooling in a synchrotron-emitting region whose temperature has approached a threshold of ${\sim}5\times 10^{11}$\,$\rm K$ \citep{1969ApJ...155L..71K}. High brightness temperatures are especially measured during flares, and indicate that VLBI cores are strongly particle-dominated. In the jet, instead, a rapid temperature drop is observed, with values which are usually in very good agreement with the equipartition limit of ${\sim}5\times 10^{10}$ $\rm K$ \citep{1994ApJ...426...51R}, i.e., with the temperature expected when the energy of the radiating particles equals the energy stored in the magnetic field.

The details of the processes taking place between the jet launching region, where the jet is a purely electromagnetic stream, and the matter-dominated cm-VLBI core are not known yet. Decisive constraints can only come from direct observational probes of the relevant spatial scales, which is where mm-VLBI comes into play.

\section{\large AGN science with mm-VLBI}
In the following part of this review we focus on the discussion of the main open questions concerning the physics of AGN jets that can be ideally addressed by the use of the mm-VLBI technique. The results presented are mainly from observations at 43\,GHz and 86\,GHz, but include some of the first findings at 230 GHz. At 43\,GHz, an important contribution comes from the blazar monitoring program run by the Boston University\footnote{https://www.bu.edu/blazars/VLBAproject.html} \citep{2016Galax...4...47J} with the VLBA. 

We start by describing the possible mechanisms giving rise to the high-energy emission and how mm-VLBI observations can help constraining its location (Sect. 4.1). From this discussion, the importance of probing the internal structure of the flow and its morphological aspects will emerge, a topic which is expanded in Sect. 4.2. We will then address the broad subject of jet formation, investigated through polarization studies (Sect. 4.3) or through the direct imaging of the jet launching region (Sect. 4.4). We will conclude with a short overview of the first scientific results and future 
goals in the study of the compact radio source in our Galaxy, Sagittarius A*, with the Event Horizon Telescope. 
\subsection{\large High-energy emission}
Since the launch of \textit{CGRO/EGRET} \citep[Compton Gamma-Ray Observatory/Energetic Gamma-Ray Experiment Telescope,][]{1993ApJS...86..629T}, \textit{AGILE} \citep[Astro-rivelatore Gamma a Immagini Leggero,][]{2008NIMPA.588...52T}, and especially of the \textit{Fermi}/{\rm LAT} \citep[Large Area Telescope,][]{2009ApJ...697.1071A}, blazars have been established as the most numerous class of $\gamma$-ray sources in the sky. To date, \textit{Fermi} has identified over 1000 blazars \citep{2015ApJS..218...23A}, a fraction of which is also detected at TeV energies\footnote{www.tevcat.uchicago.edu}. 

The most frequently invoked process to account for the high-energy emission observed in AGN is the inverse Compton (IC) scattering of soft photons by the population of relativistic electrons forming the jet. In this scenario, the electrons are giving rise to both the low-frequency synchrotron component and the high-frequency bump characterizing the spectral energy distribution (SED). 
In principle, the reservoir of soft photons can originate in various regions of the AGN. In the synchrotron self-Compton mechanism, the same synchrotron photons emitted by the jet are up-scattered to higher energies \citep{1992ApJ...397L...5M}. Concerning the ``external'' reservoirs, instead, several possibilities have been proposed. In order of increasing distance from the central engine, these include optical/UV photons from the accretion disk \citep{1993ApJ...416..458D} and from the broad line region \citep{1994ApJ...421..153S}, infra-red emission from the torus \citep{2000ApJ...545..107B}, and, on larger scales, CMB photons \citep[e.g.,][]{2001MNRAS.321L...1C}.  Alternatively, if ultra-relativistic protons are also present in the jet, the $\gamma$-ray emission may result from proton-synchrotron or from $p\gamma$ photopion production \citep{1992A&A...253L..21M, 2000NewA....5..377A}.
Especially in hadronic models, but also in the leptonic ones, extremely efficient acceleration mechanisms are required and must be at play in AGN jets.  

Statistical studies of large samples \citep[][]{2011ApJ...741...30A,2014MNRAS.441.1899F,2015MNRAS.452.1280R} have inferred the existence of significant correlations between the broad-band ($0.1$ ${\rm GeV}$ $< E < 300$  ${\rm GeV}$) \textit{Fermi} light-curves and the radio ones obtained from single-dish monitoring programs at centimeter and millimeter wavelengths (see e.g., the \textit{F-GAMMA} legacy Program\footnote{http://www3.mpifr-bonn.mpg.de/div/vlbi/fgamma/}), suggesting a common origin of the emission in the two bands. In the aforementioned studies, the radio variability is usually found to be delayed with respect to the high frequencies. Similar conclusions were reached through a VLBI analysis of the \textit{MOJAVE} sample \citep{2010ApJ...722L...7P} showing, in addition, that the radio-$\gamma$ correlations are highly significant when considering the radio properties of the core, rather than those of the jet. 

From a theoretical standpoint, the observed behavior may be conveniently explained if the activity -- triggered, for instance, by a prominent change in particle density of the plasma -- arises in a synchrotron self-absorbed region located in the innermost surroundings of the central engine. Here, highly relativistic electrons in the jet can up-scatter the optical/UV photons from the broad-line region to $\gamma$-ray energies. At energies above $1$ $\rm GeV$, this dense photon-field is also expected to be highly opaque to the emerging $\gamma$-rays through the reaction $\gamma\gamma\rightarrow e^{\pm}$ \citep[see, e.g.,][]{2009MNRAS.397..985G}, which could account for the GeV spectral break observed in several blazars \citep{2010ApJ...717L.118P}. Moreover, one of the main  arguments in favor of the ``close dissipation'' scenario is given by variability timescales as short as few hours, or even minutes \citep[e.g.,][]{2007ApJ...664L..71A, 2016ApJ...824L..20A}, which constrain the size of the emitting regions 
involved to be highly compact and have led to the conclusion that $\gamma$-rays must originate within $10^{16}$ $\rm cm$, i.e., $<<1$ $\rm pc$, from the central engine \citep{2010MNRAS.405L..94T}.  

It has also been argued, however, that assuming a direct correspondence between the inferred size of the region and its distance from the central engine may be incorrect \citep[e.g.,][]{2010arXiv1005.5551M}. While showing very rapid variability, the high-energy emission may still be produced at larger separations from the jet base if the emission region occupies only a small fraction of the jet cross-section. 

Based on the results from the Mets\"{a}hovi monitoring at 22 and 37 GHz of the \textit{EGRET}-detected blazars, \cite{1995A&A...297L..13V} were among the first to propose that, especially in quasars, the $\gamma$-ray flares are likely to originate in the millimeter-wave emitting regions \citep[see also][]{1996A&AS..120C.491V, 2003ApJ...590...95L, 2011A&A...532A.146L} rather than in the vicinity of the broad line region.
Thanks to its ability to detect detailed structural changes in the flow, mm-VLBI has now provided compelling evidence that, in fact, the high-energy events and the millimeter-wave emission are often co-spatial, corroborating physical scenarios where the $\gamma$-ray emission is produced at parsec distances from the central engine. Some of the mm-VLBI studies supporting this idea, as well as findings possibly requiring alternative interpretations, are summarized in the following.

Already before the launch of \textit{Fermi}, a systematic VLBI study of a large sample of $\gamma$-ray blazars detected by \textit{EGRET} was conducted at frequencies up to 43\,GHz by \cite{2001ApJS..134..181J}. In a large number of sources, a clear connection is observed between the occurrence of a $\gamma$-ray flare and the ejection of a new superluminal component. The $\gamma$-ray flare is found, on average, to follow the ejection, and to nearly coincide with a local maximum in the polarized radio flux density. By interpreting the newly ejected components as traveling shocks, this delay could coincide with the time required for the shock to fully develop and for the electrons to be efficiently accelerated at its forward layer. 

More recent VLBI monitoring campaigns of single objects of 43\,GHz have confirmed that, in general, a tight relation exists between the $\gamma$-ray emission and the properties of the mm-VLBI core region. In sources like 1156+295 \citep{2014MNRAS.445.1636R}, PKS 1510$-$089 \citep{2014A&A...569A..46A} and 0954+658 \citep{2014AJ....148...42M}, the $\gamma$-ray outbursts are found to be triggered by the passage of new superluminal components through the mm-VLBI core. In the case of 3C\,120 \citep{2015ApJ...808..162C} and CTA\,102 \citep{2015ApJ...813...51C}, this trend is confirmed, but only when the newly ejected features are traveling in a direction closer to the observer's line of sight. Indeed, a close connection between the jet orientation and the occurrence of $\gamma$-ray flares is evident at a statistical level in the \textit{MOJAVE} survey \citep{2009A&A...507L..33P}, since the class of $\gamma$-ray loud jets is found to be oriented at smaller viewing angles with respect to the 
general 
radio-loud population. Flaring events triggered by the 
ejection of new components are often accompanied by increased activity in the optical and X-rays, and by systematic rotations of the optical electric vector position angle \citep[e.g.,][]{2008Natur.452..966M, 2010ApJ...715..362J, 2010ApJ...710L.126M}.

\begin{figure}[!h]
\centering
\includegraphics[width=0.9\textwidth]{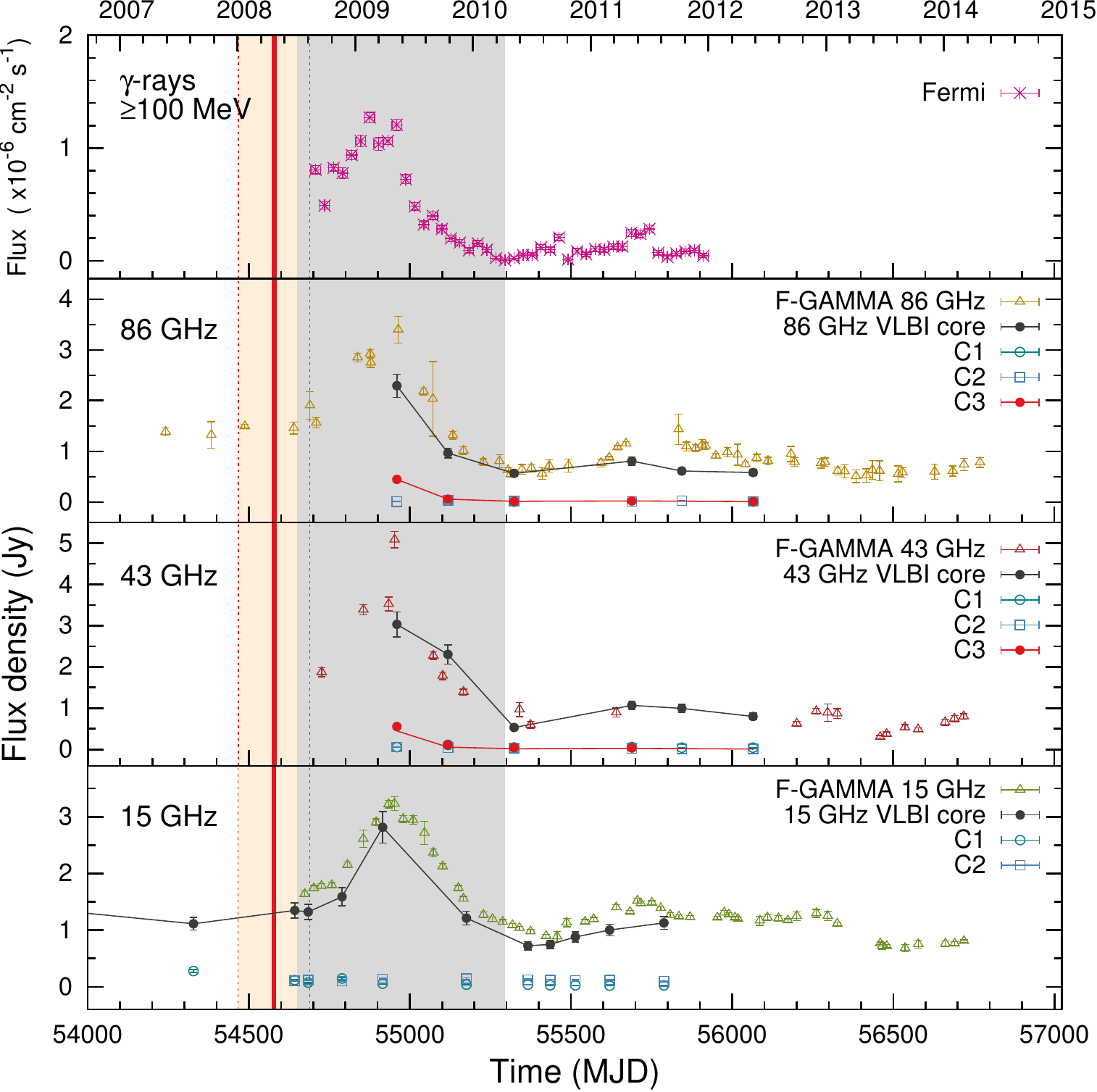}
 \caption{Light curves of the blazar PKS\,1502+106, which showed a very prominent $\gamma$-ray flare in 2009 \citep{2016A&A...590A..48K}. From top to bottom: 1) monthly binned {\it Fermi}/LAT $\gamma$-ray light curve at energies $E>100$ $\rm MeV$; 2) F-GAMMA single-dish radio light curve at 86\,GHz, core and component light curves from the VLBI flux density decomposition; 3) same as above, at 43\,GHz; 4) same as above, at 15 GHz. A newly ejected VLBI knot, labeled C3, is visible at 43\,GHz and 86\,GHz. Its ejection time, designated by the red solid line and estimated based on the VLBI kinematic analysis, follows the onset of the $\gamma$-ray flare. The latter originates ${\sim}2$ parsecs away from the black hole, downstream of the broad line region \citep{2016A&A...586A..60K}. (Image reproduced with permission from \cite{2016A&A...590A..48K}, copyright by ESO)}
\end{figure}

In several cases, the mm-core where the high-energy emission is produced has been identified as a standing shock \citep[e.g.,][]{2007ApJ...659L.107D, 2008Natur.452..966M, 2010ApJ...710L.126M}, possibly marking the end of the acceleration and collimation zone. Standing shocks playing a major role in the high-energy flaring events have been pinpointed with mm-VLBI also downstream in the jet.
In OJ\,287, a secondary standing shock at a distance of 14 parsecs from the innermost stationary feature was proposed to give rise to a prominent $\gamma$-ray flare as it was crossed by a turbulent moving blob \citep{2011ApJ...726L..13A}. In 3C\,345, \cite{2012A&A...537A..70S} showed that $\gamma$-rays may be produced at multiple locations over an extended region of 23 parsecs, including the proximity of a stationary feature observed ${\sim}10$ parsecs away from the core.

While the physical processes triggering the activity may vary from flare to flare, all of the above results point toward a ``far dissipation'' scenario. Although, as we discussed in Sect. 3, observing at millimeter wavelengths enables us to unveil emission regions closer to the jet base, the millimeter core is still located at parsec distances from the black hole, at least in the brightest blazars \citep[e.g.,][]{2008ApJ...675...71S, 2015A&A...576A..43F}.
The inferred co-spatiality of the low and high-energy emission implies that efficient particle acceleration takes place far beyond the broad-line region. 
At the distances determined by mm-VLBI studies, the most likely mechanisms giving rise to the high-energy emission are then the synchrotron self-Compton and the external inverse Compton of infra-red photons from the torus.

This conclusion may still be valid when the $\gamma$-ray activity is found to precede the variations in the radio band \citep[see, e.g.,][]{2014A&A...571L...2R, 2016A&A...590A..48K, 2017MNRAS.468.4478L}. For instance, through a dedicated VLBI campaign at 43\,GHz and 86\,GHz of the high-redshift blazar PKS\,1502+106, \cite{2016A&A...590A..48K} determined that the high-energy emission originates few parsecs upstream of the mm-core (Fig. 6). With the combined effort of single-dish and VLBI studies, an absolute distance of ${\sim}1.9$\,$\rm pc$ was inferred for the location of the $\gamma$-ray activity with respect to the black hole \citep{2016A&A...586A..60K}. Since the broad line region is expected to extend on much smaller scales in this object \citep[${\sim}0.1$\,$\rm pc$,][]{2010ApJ...710..810A}, an external seed photon field is most likely provided by the infrared torus also in this case. 

Next to the described findings, a number of cases which may require more complex interpretations have been reported in the literature.
`Orphan'' $\gamma$-ray flares, with no structural change on parsec scales, have been frequently observed, e.g., in Mrk\,421 \citep{2014A&A...571A..54L} and CTA\,102 \citep{2015ApJ...813...51C}. In the misaligned object 3C\,84, the variability in the radio and at the very high $\gamma$-ray energies ($E>10$ ${\rm GeV}$) appears to be uncorrelated \citep{2012ApJ...746..140S, 2012MNRAS.423L.122N}.  Particularly for the TeV BL Lacs and for radio galaxies, whose Doppler factors inferred from VLBI kinematic analyses are found to be too low to account for the observed energetic processes, a simple ``one-zone model'' for the high-energy production may be not appropriate. For these objects it was suggested \citep{2005A&A...432..401G} that the inverse Compton emission can be enhanced if the jet is structured transversely, e.g., it is formed by a fast central spine and a slower outer sheath. Under these 
conditions, the electrons of one component can up-scatter the beamed photons of the other, so that the low and the high-energy emissions do not necessarily originate in the same region, and can show uncorrelated variability. A similar mechanism was theorized by \cite{2003ApJ...594L..27G} for decelerating jets, i.e., those characterized by a radial velocity gradient.  43\,GHz VLBI observations of 3C\,84 have revealed that this jet is in fact characterized by a transverse structure of the spine-sheath kind \citep[Fig. 7,][]{2014ApJ...785...53N}. The uncorrelated radio-$\gamma$ variability, the observed transverse stratification, and modeling of the SED favor a ``two-zone model'' for the high-energy production in this source \citep{2014MNRAS.443.1224T}. 

Transversely stratified jets in BL Lacs have been proposed \citep{2014ApJ...793L..18T} as possible sources of the neutrinos detected by IceCube in the (0.1-1) PeV range \citep{2013PhRvL.111b1103A}. These detections have opened new frontiers for the study of the high-energy emission in AGN, since several recent studies have identified a number of blazars as possible counterparts of the IceCube signals \citep[e.g.,][]{2016NatPh..12..807K,2016MNRAS.457.3582P}.
Future mm-VLBI analysis of the structural characteristics of these candidates may provide further clues for the correctness of the proposed associations.

\subsection{\large Internal structure and morphology}
As evident from the previous section, probing the morphology and the internal structure of the innermost regions of AGN jets is essential 
for truly understanding the physical processes occurring in the plasma flow. Analyzing the jet transverse structure is not only important for the correct modeling of the high-energy emission, as explained, but also for probing the development of plasma instabilities, which can affect crucially the jet propagation on larger scales. 

In order to obtain meaningful insights into the internal jet structure, it is necessary to resolve the flow in the transverse direction. This condition is usually not met in centimeter VLBI observations, especially in the most compact regions at the jet base. Interferometric observations with enhanced resolution, like space or mm-VLBI, are instead adequate for achieving such a goal in several nearby objects. 

A powerful tool for testing the development of instabilities is the analysis of the jet ridge line, i.e., of the evolution of the location of the peak of emission in the transverse direction as a function of distance from the core. In the case of the strong jet in S5\,0836+710 \citep{2012ApJ...749...55P}, VLBI observations up to 43\,GHz have shown that the ridge line is likely tracing a pattern of helically twisted pressure maxima due to the formation of Kelvin--Helmoltz instabilities. A similar conclusion was reached in the analysis of 3C\,273 through space VLBI observations at 5 GHz \citep{2001Sci...294..128L}. In this case, a double helical ridge line is observed, the jet being bright at its limbs and dimmer close to the central axis. 

\begin{figure}
\centering
 \includegraphics[width=0.55\textwidth]{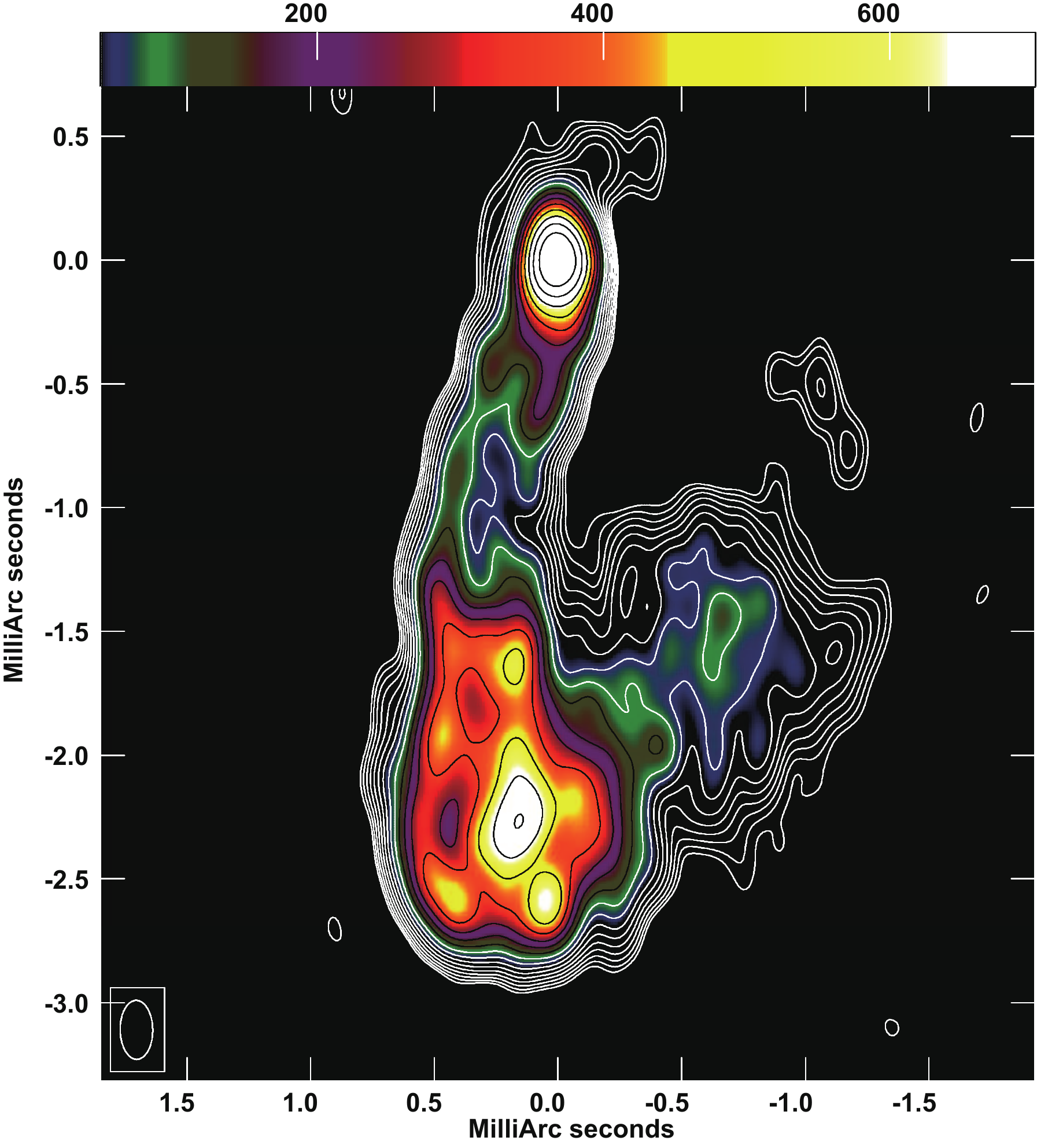}
 \caption{The limb-brightened jet structure of the peculiar radio galaxy 3C\,84 as revealed by VLBI observations at 43\,GHz. (Image reproduced with permission from \cite{2014ApJ...785...53N}, copyright by AAS)}
\centering
\vspace{0.2cm}
 \includegraphics[width=0.85\textwidth]{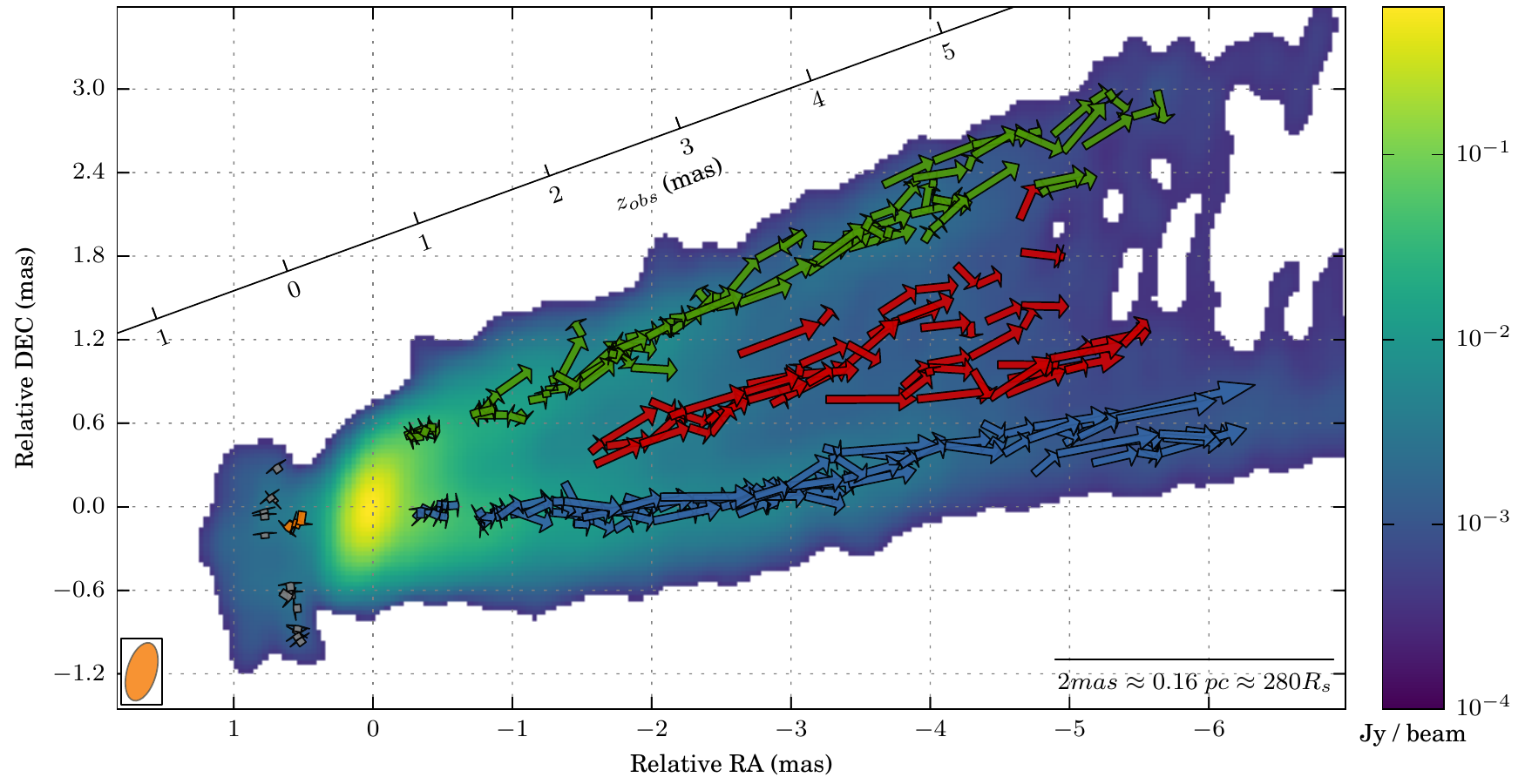}
 \caption{Velocity structure of the jet in M\,87 derived from the analysis of a multiepoch VLBI data set at 43\,GHz. Motion is detected along three main filaments, two outer limbs and a central body. Two overlapping velocity components can be identified: a mildly relativistic one and a faster streamline with a Lorentz factor of 2.5. (Image reproduced with permission from \cite{2016A&A...595A..54M} copyright by ESO)}
\end{figure}

The class formed by the so-called ``structured jets'', i.e., those characterized by a transverse gradient in the intensity, in the velocity, or in the polarization properties (see Sect. 4.3), is getting more and more numerous as the resolution of the images improves.  
Several limb-brightened jets could recently be imaged through mm-VLBI, e.g., in 3C\,84 \citep[][Fig. 7]{2014ApJ...785...53N}, in Mrk\,501 \citep{2008A&A...488..905G}, in Mrk 421 \citep{2010ApJ...723.1150P}, in M\,87 \citep{1999Natur.401..891J, 2016Galax...4...46W, 2016A&A...595A..54M}, and in Cygnus\,A \citep{2016A&A...585A..33B, 2016A&A...588L...9B}. In the latter source, the stratification is visible not only in the approaching jet, but also in the counter-jet, and a similar feature is hinted in the weak counter-jet of M\,87. The limb-brightening characterizing these jets could reflect the presence of instability patterns, as for 3C\,273, or it may be a direct result of the jet formation mechanism. This second scenario appears especially plausible for M\,87, where the double ridge line is visible already at few tens of Schwarzschild radii from the central engine.
The intensity gradient can, in this case, be explained in two ways: 1) the jet emissivity is intrinsically lower closer to the jet axis or 2) the jet emissivity is apparently lower close to the jet axis due to the existence of a spine-sheath velocity gradient. In fact, limb-brightening will naturally arise in sufficiently misaligned jets due to the differential boosting of each filament or to the de-boosting of the fast central spine. A spine-sheath velocity structure has been long proposed for relativistic jets \citep{1989MNRAS.237..411S}, and is compatible with findings from recent kinematic studies in Cygnus\,A \citep{2016A&A...585A..33B} and M\,87 \citep{2016A&A...595A..54M}. In particular, \cite{2016A&A...595A..54M} were able to obtain a detailed 2-D velocity field in the inner regions of M\,87, clearly showing the displacement of features along three main filaments of the flow, two outer limbs and a central body (Fig. 8). Two speed components, which appear to overlap across the jet, are detected: a 
mildly relativistic speed likely associated with an outer disk wind or with an instability pattern and a faster streamline featuring a Lorentz factor of ${\sim}2.5$.

These results point out the importance of resolving transversely the plasma flow in order to better understand the fundamental processes involved in its formation, a subject which will be more widely debated in the final part of this article. 
Another morphological aspect which is emerging as a characteristic feature of jets in mm-VLBI images, and that may be also intimately related to the launching mechanism, is the large apparent bending of emission features in the immediate surroundings of the core, also know as ``jet wobbling''. Changes in the direction of the jet axis appear to be more dramatic close to the jet apex than on the scales probed by cm-VLBI observations. Strongly bending features, which are often stationary, are seen on sub-milliarcsecond scales in objects like Mrk\,501 \citep{2016A&A...586A.113K}, OJ\,287 \citep{2017A&A...597A..80H}, 3C\,279 \citep{2013ApJ...772...13L}, NRAO\,150 \citep{2007A&A...476L..17A}, CTA\,102 \citep{2013A&A...551A..32F}, and other sources \citep[e.g.,][]{1998A&AS..131..451R}.
At present, no clear explanation exists for their appearance. In some objects, the wobbling appears to be characterized by a periodicity, as suggested for the blazar 4C\,+12.50 \citep{2003ApJ...584..135L}. In this case, a regular precession of the accretion disk, either induced by a binary black hole system or by the presence of some other massive object located in the nuclear regions, appears as a viable explanation for the periodic change of direction. In most of the cases, however, the jet bending is rather erratic. In OJ\,287, \cite{2012ApJ...747...63A} exclude the existence of well-defined periodicity of the inner-jet swing \citep[the 12-year periodic precession proposed by][for larger scales]{2004ApJ...608..149T}, and rather favor phenomena of asymmetric injection of plasma due to turbulence in the accretion disk or of instabilities in the innermost jet. If the magnetic field is still dynamically important on the scales probed by mm-VLBI, current-driven instabilities \citep{1981ApJ...247..792B} are 
expected to manifest. Among them, the kink instability is potentially the most disruptive \citep{2009MNRAS.394L.126M, 2010MNRAS.402....7M}, and was recently suggested to play a critical role for the origin of the FR\,I/FR\,II dichotomy in radio galaxies \citep{2016MNRAS.461L..46T}.

\subsection{\large Polarization and magnetic field topology}
Polarimetric studies can be a powerful tool for deriving fundamental constraints on the jet physics, as well as on the properties of the jet environment. 
The interpretation of polarization data requires, however, particular caution. The likely co-existence of relativistic, projection and other effects makes it challenging to reconstruct the intrinsic orientation and strength of the magnetic fields \citep[see e.g.,][]{2005MNRAS.360..869L}. Moreover, the very nature of these fields is still highly debated. In particular, it is unclear whether the observed polarization has to be ascribed to the local ordering, e.g., due to compression in a shock, of an otherwise random ambient magnetic field, or to the presence of a large-scale, ordered field permeating the plasma flow. While a certain degree of ambiguity is likely inevitable when investigating this subject, we discuss here how VLBI observations at millimeter wavelengths enable many of the uncertainties to be reduced.

From synchrotron theory, the predicted fractional linear polarization for an ensemble of relativistic electrons moving in a uniform magnetic field is 70\%--75\%, in the case of optically thin emission, and 10\%--12\% for optically thick emission \citep{1970ranp.book.....P}. However, both single-dish radio surveys \citep[e.g.,][]{1985ApJS...59..513A, 1992ApJ...399...16A} and VLBI studies of large samples \citep[e.g.,][]{2003ApJ...589..733P, 2005AJ....130.1389L, 2007ApJ...658..203H} have inferred much lower degrees than expected on this theoretical basis. Typical values measured in the proximity of the VLBI core are below 5\%, while there is a trend of increasing fractional polarization, up to tens of percent, at larger distances from the jet base \citep[e.g.,][]{2005AJ....130.1389L}. 

These results have often been interpreted as evidence for the random nature of the magnetic field \citep[see][and references therein]{2005ApJ...621..635H}.
Under the assumption that large part of the jet emission originates in shocked regions, polarization degrees of 20--30\%, or higher, can be explained in the framework of the shock-in-jet model \citep{1985ApJ...298..114M, 1985ApJ...298..301H}, as a consequence of the compression and amplification of the component of the magnetic field oriented parallel to the shock front. In the case of a transverse shock, the electric vector position angle (EVPA) is then expected to be oriented in the direction parallel to the jet axis, a configuration which is frequently found to characterize the compact jet knots in polarimetric VLBI images, especially in BL Lac objects \citep{2000MNRAS.319.1109G}.
Enhanced polarization can also result from the formation of oblique and conical shocks \citep[][and references therein]{1990ApJ...350..536C} or from the shear of the magnetic field lines at the boundaries between the jet and the ambient medium \citep{1980MNRAS.193..439L}.

Although a largely random field can be adequate for reproducing the observed properties of AGN jets, its assumption poses some difficulties for theoretical models describing the launching, acceleration, and collimation of the flow. In fact, most of these models require the existence of an ordered, large scale field at the base of the jet \citep[e.g.,][and references therein]{2010LNP...794..233S}. If this prediction is correct, the aforementioned observational results can then be explained in two ways. A first possibility is that the field gets disrupted and tangled, e.g., owing to magnetic instabilities \citep{2006A&A...450..887G}, already before the jet becomes visible at the VLBI core. This scenario is plausible especially in the case of the most highly boosted blazars, where, as we said, the mm-VLBI core appears to be located quite far from the central engine, at a distance of $10^4 -10^5$ Schwarzschild radii
\citep[e.g.,][]{2008Natur.452..966M, 2015A&A...576A..43F}. Alternatively, a large scale field may be preserved on VLBI scales, and the radio emission may be intrinsically highly polarized, but a strong depolarization occurs as the radiation propagates to the observer. In general, depolarization results 
as a consequence of the integration of polarized emission from multiple emission regions and/or along the line of sight. In relativistic jets, a decrease of the net polarization could be observed in the following cases: 
\begin{itemize}
\item \textit{Line of sight integration through opaque material}\\
\indent  Assuming that the VLBI core actually marks, at a given frequency, the transition region to the optically thin regime, then at least part of its emission is expected to be opaque. The line of sight integration through a partially optically thick material could be responsible for the low net polarization observed at the core location. This effect may apply especially to sources which are seen at a very small viewing angles.
\item \textit{Beam depolarization}\\
\indent VLBI observations are characterized by a finite resolution. If the observing beam is not negligible when compared to the size of the coherent polarization structure of the source, depolarization may result from the averaging of polarized emission inside the interferometer beam.
\item \textit{Faraday depolarization}\\
\indent In addition to the previous effects, further depolarization could arise as a consequence of Faraday rotation \citep{1966MNRAS.133...67B}, i.e., the rotation of the polarization plane of an electromagnetic wave while it propagates through a magnetized plasma. The rotation measure $RM$ quantifies, at a given wavelength $\lambda$, the difference between the intrinsic polarization angle $\chi_0$ and the observed one $\chi$. In the simplest case of a homogeneous medium, $RM$ is proportional to the integral over the line of sight from the source to the observer of the density of charges in the medium $n_e$ times the component of the magnetic field along the line of sight $B_{\parallel}$:
\begin{equation}
\chi=RM \lambda^2 + \chi_0, \quad\quad RM \propto \int{n_e B_{\parallel} dl}.
\end{equation}
The presence of Faraday screens can cause a reduction of the total polarization whenever the emission experiences differential rotations while propagating to the observer. This can happen either when the Faraday rotating material is mixed with the emitting plasma (internal depolarization) or when the radiation crosses external inhomogeneous media (external depolarization). It is important to note that the strongest rotations are expected to be induced by thermal electrons, whose presence is likely abundant in the jet environment. Dense thermal media which can potentially act as Faraday screens include both orbiting clouds in the broad- or narrow-line region and non-relativistic material embedding the relativistic jet beam (disk winds).
Multi-frequency cm-VLBI studies have indeed inferred the presence of such media from the high rotation measures (up to thousands of $\rm rad/m^2$) detected in several blazars \citep[e.g.,][]{1998ApJ...506..637T, 2004ApJ...612..749Z, 2012AJ....144..105H}. The effect appears to be especially strong at shorter distances from the central engine, indicating a decreasing strenght of the magnetic field with distance and/or a decreasing density of the thermal electrons. The latter is in agreement with the expectations from the unified scheme for AGN that most of the dense gas surrounding the accretion disk is concentrated in the central parsecs. 
\end{itemize}

Ultimately, while a large scale field may still permeate the plasma flow on small VLBI scales, these very scales are also the most heavily obscured, synchrotron self-absorbed and unresolved. VLBI at millimeter wavelengths is then the optimal technique when aiming at revealing the intrinsic strength and orientation of the magnetic fields in the jet. Due to its dependence on the square of the wavelength (eq. 7), Faraday rotation is expected to be diminished, and so is the depolarization arising from it. Moreover, being able to penetrate the jet base closer to the central engine, mm-VLBI increases the possibility to image -- with enhanced resolution -- those regions potentially characterized by dynamically important magnetic fields.

A clear example of the potential of mm-VLBI observations is provided, in this context, by the study of BL Lac presented by \cite{2008Natur.452..966M}. 
At 43\,GHz, polarized emission could be detected from a region upstream of the VLBI core (Fig. 9). The disturbance injected at the jet base propagates downstream, causing a double flare at optical, X-rays and TeV energies. The second flare is also observed in radio when the disturbance crosses the VLBI core, which is interpreted as a standing shock. The optical and radio polarization properties suggest the upstream region to be part of the acceleration and collimation zone of this jet. In particular, the smooth rotation of the optical polarization vector by $\sim240^{\circ}$ observed over few days and its final alignment with the radio polarization vector, parallel to the jet axis, support the presence of a large scale magnetic field with helical geometry which then becomes mostly turbulent close to the standing shock. In this scenario, consistent with the findings from RadioAstron at 22 GHz \citep{2016ApJ...817...96G} and similarly suggested for PKS 1510$-$089 \citep{2010ApJ...710L.126M}, the VLBI core is 
the 
location where the plasma has attained its terminal speed, and possibly equipartition of the magnetic and particle energy densities.  
\begin{figure}[!t]
\centering
 \includegraphics[width=0.95\textwidth]{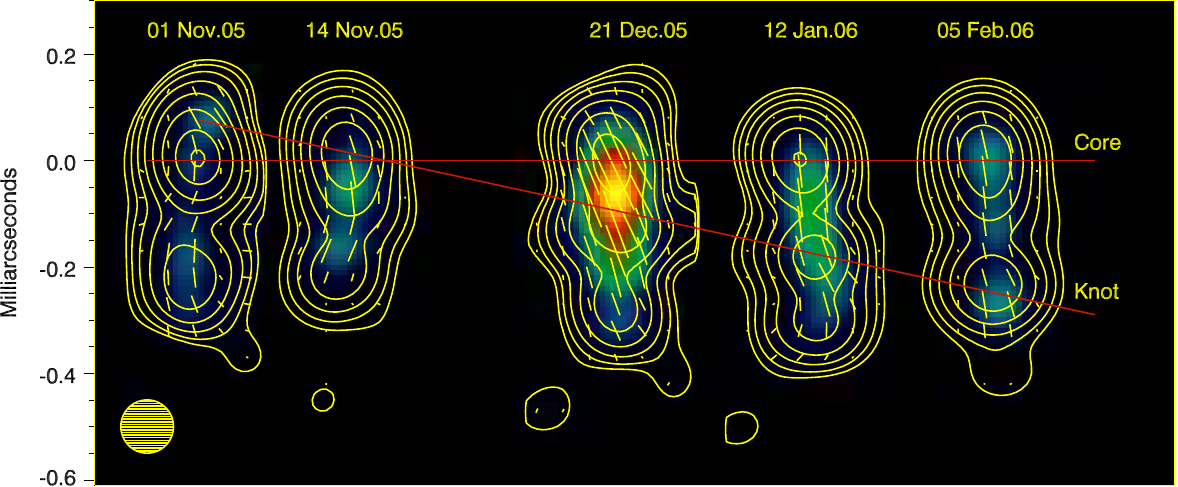}
 \caption{VLBI polarimetric images of BL Lac at 43\,GHz. A feature emerging in the region upstream of the core is visible in the first three maps. The optical and radio polarization properties suggest this region to be part of the jet acceleration and collimation zone. (Image reproduced with permission from \cite{2008Natur.452..966M}, copyright by NPG)}
\end{figure}

In the picture proposed by \cite{2008Natur.452..966M}, the helical structure of the magnetic field is already disrupted at the mm-VLBI core, located at a distance of ${\sim}10^4$\,$R_S$ from the jet apex. Several other authors, however, have suggested that the helical geometry may be preserved on much larger scales. The existence of a parsec-scale helical field has been proposed for NRAO\,150 \citep{2014A&A...566A..26M} for explaining the high speed rotation of the emission regions. In the polarimetric analysis of this object performed up to 86\,GHz frequencies, a toroidal magnetic field is observed, although the polarization degree is quite low due to the almost perfectly face-on orientation of this jet.
In \cite{2013MNRAS.436.3341Z}, the nature of a broad arc-like feature in the inner regions of 3C\,454.3 was investigated through multi-frequency, polarimetric VLBI imaging between 5 GHz and 86\,GHz. Transverse gradients across the arc, which is highly polarized, are observed both in the polarization degree and in the apparent magnetic vector position angle, which varies by more than $90^{\circ}$. Moreover, the rotation measure shows a sign reversal across the feature. These properties can be best reproduced by assuming the presence of a large-scale helical field. If confirmed, this result has the remarkable implication that an ordered field exists at distance of ${\sim}10^7$ $R_{\rm S}$, i.e., much beyond the expected extension of the acceleration and collimation region. Transverse rotation measure gradients have been observed in several other objects on parsec scales \citep[e.g.,][]{2002PASJ...54L..39A, 2004MNRAS.351L..89G, 2009MNRAS.393..429O}, and often interpreted as a characteristic signature of helical 
fields. The analysis of RM transverse gradients can provide important insights for the investigation of this subject. However, as pointed out by \cite{2010ApJ...722L.183T}, spurious results can arise as a consequence of limited transverse resolution in the VLBI images. Therefore, it will be important in the future to verify the aforementioned results through VLBI studies at the highest possible resolution. 

Polarimetric imaging at frequencies higher than 43\,GHz has been limited in the past due to the non-standard calibration procedures required. 
Over the past few years, a pipeline for the calibration of full-polarization, global VLBI data at 86\,GHz has been developed, and the feasibility of the technique has been demonstrated for the case of the blazar 3C\,345 \citep{2012A&A...542A.107M}. Recently, a polarimetric image of M\,87 has also been produced at this frequency \citep{2016ApJ...817..131H}, revealing the presence of a highly polarized feature in the innermost regions of the jet. 
Concerning the earlier experiments, 86\,GHz polarization maps had been obtained for mainly three objects: 3C\,120 \citep{1999ApJ...521L..29G}, 3C\,273 \citep{2001ApJ...553L..31A, 2005ApJ...633L..85A}, and 3C\,279 \citep{2001ApJ...553L..31A}. No polarization was detected for the cores of 3C\,120 and 3C\,273, while a weakly polarized core was observed in 3C\,379. At larger distances from the central engine, instead, highly polarized features emerge in all the three cases, with the magnetic field oriented parallel to the jet axis in 3C\,120 and 3C\,273, and perpendicular in 3C\,279. If the absence of polarization in the core is due to Faraday depolarization only, extremely high rotation measures, larger than $90000$\,$\rm rad$ $\rm m^{-2}$ in the case of 3C\,273, are inferred. While, on the one hand, several alternative explanations exists for the low polarization of VLBI cores, such high rotation measures are not incompatible with the tremendous opacities that can be expected at the jet base. Through recent 
ALMA observations of the high-redshift blazar PKS 1330$-$211 at frequencies up to 300 GHz (1 THz in the source frame), rotation measures of the order or ${\sim}10^8$  $\rm rad$ $\rm m^{-2}$ have been derived \citep{2015Sci...348..311M}. These values are consistent with the presence of dynamically important magnetic fields at the jet base, with strengths of tens of Gauss and potentially much higher. The importance of this and other similar results is further commented in the following section. 

\subsection{\large Launching, acceleration and collimation}
VLBI studies at centimeter wavelengths, as well as radio interferometric analyses probing larger scales, have collected substantial observational evidence of one basic fact: jets are self-similar over many orders of magnitude in length. This self-similarity is expected to break close to the black hole, where the jet should be dominated by the magnetic energy driving it. Among the proposed mechanisms for jet formation, many of which involve a purely hydrodynamical launching \citep[][]{1974MNRAS.169..395B}, the magnetic launching model \citep[see e.g.,][]{2001Sci...291...84M} appears as the most viable option according to numerical simulations \citep[e.g.,][and references therein]{2015ASSL..414...45T}, and is currently largely favored. 
In this model the jet properties scale trivially with the black hole mass, conveniently explaining the similarities between galactic and extragalactic jets. 

Although there is a general agreement on the basic principles of the magnetic launching mechanism, 
reproducing the extreme properties observed in relativistic jets, such as opening angles smaller than 1 degree and bulk Lorentz factors as high as 50, is still among the most challenging tasks for theorists. This is firstly due to the inherent complexity of plasma physics in the immediate surrounding of super-massive black holes, whose description requires a full GRMHD (General Relativistic Magneto-HydroDynamic) formalism. In the past 20 years, important steps forward have been made by implementing three-dimensional GRMHD numerical simulations \citep[][]{1999ApJ...522..727K, 2006MNRAS.368.1561M}. These have shown that AGN jets can be efficiently powered by the rotational energy of the compact central object, extracted through strong, large scale magnetic fields. 
The power source can either be the accretion disk, as described in the work of \citep{1982MNRAS.199..883B}, or a spinning super-massive black hole, as originally proposed by \cite{1977MNRAS.179..433B}. 

At their very base, such magnetically-driven jets are likely to propagate in the form of pure Poynting flux, which then gets gradually converted into kinetic flux and, for a small percentage, into radiation. Based on theoretical models \citep{2004ApJ...605..656V, 2009ApJ...698.1570L} and numerical simulations \citep{2006MNRAS.368.1561M, 2007MNRAS.380...51K}, the predicted scales for the mechanisms of acceleration and collimation to take place span an interval between few and $10^3-10^5$ Schwarzschild radii, corresponding to sub-parsec or parsec scales (see Fig. 5). In this region, the magnetic field is expected to be dynamically important and well ordered, with a geometry gradually evolving from purely poloidal to helical, or possibly to purely toroidal. 

In the recent years, mm-VLBI observations have been able to probe these previously unexplored scales following different approaches. In addition to the analysis of the polarization properties, discussed in the previous section, important observational tests could be provided either through statistical studies in survey experiments or through a direct and detailed imaging of single, optimal targets. 

The most recent surveys conducted at 86\,GHz \citep{2000A&A...364..391L, 2008AJ....136..159L, 2016cosp...41E.710G} enabled to model the observed distribution of brightness temperature, and to investigate its dependence as a function of distance from the central engine. A significant trend has emerged, indicating that the brightness temperature measured at the location of the 86\,GHz core, peaking at $T_{\rm b}\sim10^{11}$\,$\rm K$ \citep{2008AJ....136..159L}, is lower than the one measured at lower frequencies. 
This behavior can be best explained by assuming that plasma acceleration is still taking place in the region between the VLBI cores \citep{2016ApJ...826..135L}, a result which confirms the theoretical prediction that the acceleration zone extends on parsec scales.
These findings, together with many of the previously discussed polarization and variability studies \citep[e.g.][]{2008Natur.452..966M, 2010ApJ...710L.126M}, collocate the mm-core in a crucial position: in blazars the mm-core marks the transition region between the magnetic-dominated and the kinetic-dominated regimes.

One implication of this scenario is that, in most sources, the bulk of the MHD processes driving the outflow may occur in the ``invisible'' jet, comprised between the supermassive black hole and the mm-core. There exists, however, a limited sample of objects where a high resolution imaging of the true jet base can be obtained and MHD models for jet formation can be tested in detail. The list of suitable targets mainly includes nearby misaligned jets (Table 2) where, for reasons which are not fully clarified \citep[][]{2011Natur.477..164M} but are most likely related to the reduced impact of projection and relativistic effects, the VLBI core appears to be located much closer to the central black hole than in blazars. For instance, phase-referencing VLBI observations of M\,87 \citep{2011Natur.477..185H} have determined a shift of only 14$-$23 $R_{\rm S}$ of the 43\,GHz core with respect to the central engine, while an upper limit of 100 $R_{\rm S}$ could be inferred for NGC\,1052 at 86\,
GHz \citep{2016A&A...593A..47B}. Table 2 indicates that the resolution provided by VLBI at 86\,GHz is sufficient for probing the acceleration and collimation region in all of the listed objects, while the mechanisms involved in the jet launching, expected to take place on scales $<100$\,$R_{\rm S}$ (Sect. 3), can be currently best studied in M\,87. The latter is the optimal target in terms of spatial resolution, although the intermediate orientation of its jet \citep[$\theta\sim17^{\circ}$,][]{2016A&A...595A..54M} introduces larger uncertainties on the intrinsic parameters than in true radio galaxies like Cygnus\,A  \citep[$\theta{\sim}75^{\circ}$,][]{2016A&A...585A..33B} or NGC\,1052 \citep[$64^{\circ}<\theta<87^{\circ}$,][]{2016A&A...593A..47B}.  

\begin{table}
\centering
\caption{List of nearby, misaligned objects which are well suited for mm-VLBI observations aimed at imaging the jet formation region. Col. 1: Source name. Col. 2: Luminosity distance$^*$ $D_{\rm L}$. Col. 3: Logarithm of the black hole mass $\rm M_{BH}$, given in units of solar masses $\rm M_{\odot}$. Ref: a) \cite{2011ApJ...729..119G} - b) \cite{2003MNRAS.342..861T} - c) \cite{2013MNRAS.429.2315S} - d) \cite{2002ApJ...579..530W} - e) \cite{2010PASA...27..449N}. Col. 4: Linear size in units of Schwarzschild radii probed by VLBI at 86\,GHz. The objects are listed in order of decreasing spatial resolution.}
\label{my-label}
\begin{tabular}{lllll}
\hline\noalign{\smallskip}
Source      & $D_{\rm L}$ $[\rm Mpc]$&Log$\rm(M_{BH}[\rm M_{\odot}])$& Linear size of 50$\mu$as [$R_{\rm S}$] &  \\ 
\noalign{\smallskip}\hline\noalign{\smallskip}
M\,87       & 19                             & 9.8$^a$                                & 8                              &  \\ 
Cygnus\,A   & 252                            & 9.4$^b$                                & 229                            &  \\ 
3C\,84      & 77                             & 8.9$^c$                                & 235                            &  \\ 
NGC\,1052   & 22                             & 8.2$^d$                                & 354                            &  \\ 
Centaurus\,A& 8                              & 7.7$^e$                                & 397                            &  \\ 
\noalign{\smallskip}\hline
\end{tabular}
\\
\flushleft
{\scriptsize $^*$Assuming a $\Lambda$CDM cosmology with $H_0$ = 69.6 $\rm h^{-1}$\,$\rm km$ $\rm s^{-1}$\,$\rm Mpc^{-1}$, $\Omega_{\rm M}$ = 0.286, $\Omega_{\rm \Lambda}$ = 0.714.}
\end{table}

In M\,87, the jet formation region has been investigated in numerous studies. The collimation zone is well resolved already at 43\,GHz; on sub-parsec scales, the limb-brightened flow appears much broader than on parsec scales, expanding with an opening angle of ${\sim}60^{\circ}$ \citep{1999Natur.401..891J, 2004AJ....127..119L}. Deeper imaging at 86\,GHz \citep{2016ApJ...817..131H} reveals an even broader jet base with apparent opening angle of ${\sim}100^{\circ}$, indicating that the jet starts poorly collimated and gets rapidly focused farther downstream. These findings are confirmed by the analysis of a more recent 86\,GHz image, obtained after stacking five different maps \citep[Fig. 10,][]{2016Galax...4...39K}. The stacking method is especially effective when aiming at recovering the full jet cross-section and at investigating with better fidelity the jet expansion. Both studies performed at 86\,GHz show that the jet has a parabolic shape on scales of ${\sim}100$\,$\rm R_S$. Using a rich, multi-
frequency VLBI data set, \cite{2012ApJ...745L..28A} inferred that this shape is preserved also on larger scales, up to approximately ${\sim}10^5$\,$\rm R_S$. Beyond that, however, the flow appears to be freely expanding, i.e., it has a conical shape.  According to theory, acceleration through magnetohydrodynamic processes is inefficient in conical flows, and can extend on large scales only if the plasma undergoes differential collimation \citep[see e.g.,][]{2012rjag.book...81K}. The observed transition from a parabolic to a conical shape may then signal a crucial change in the physical conditions of the plasma, marking the termination of the acceleration zone. Interestingly, the transition occurs in the vicinity of the Bondi radius (where a change in the ambient pressure gradient is expected) and of a stationary feature in the HST-1 complex, a bright region where superluminal speeds of $4c-6c$ have been measured \citep{1999ApJ...520..621B}. Then, a natural question to ask is: does the flow show acceleration 
in the parabolic region comprised between its base and HST-1? Previous kinematic studies \citep{1995ApJ...447..582B, 2007ApJ...668L..27K, 2007ApJ...660..200L, 2014ApJ...781L...2A} have yielded contrasting results, likely due to the poor sampling and to the intrinsic difficulty in identifying moving features in a jet that lacks well defined spots, and is transversely stratified. The method recently employed by \cite{2016A&A...595A..54M} enabled a detailed two-dimensional kinematic structure to be inferred, and therefore appears as the most appropriate in the case of M\,87 (Fig. 8). In this work, accelerating features are detected on scales from $10^2$ to $10^4$\,$R_{\rm S}$; the speeds, which are mildly relativistic, increase faster up to distances of $10^3$\,$R_{\rm S}$, while a milder gradient is observed further downstream. These properties could be well interpreted in the framework of MHD acceleration and collimation: the bulk of the acceleration occurs within thousands of Schwarzschild radii, and is 
followed by a regime of saturation of the Poynting flux conversion which may extend up to HST-1. 

\begin{figure}
\centering
 \includegraphics[width=0.75\textwidth]{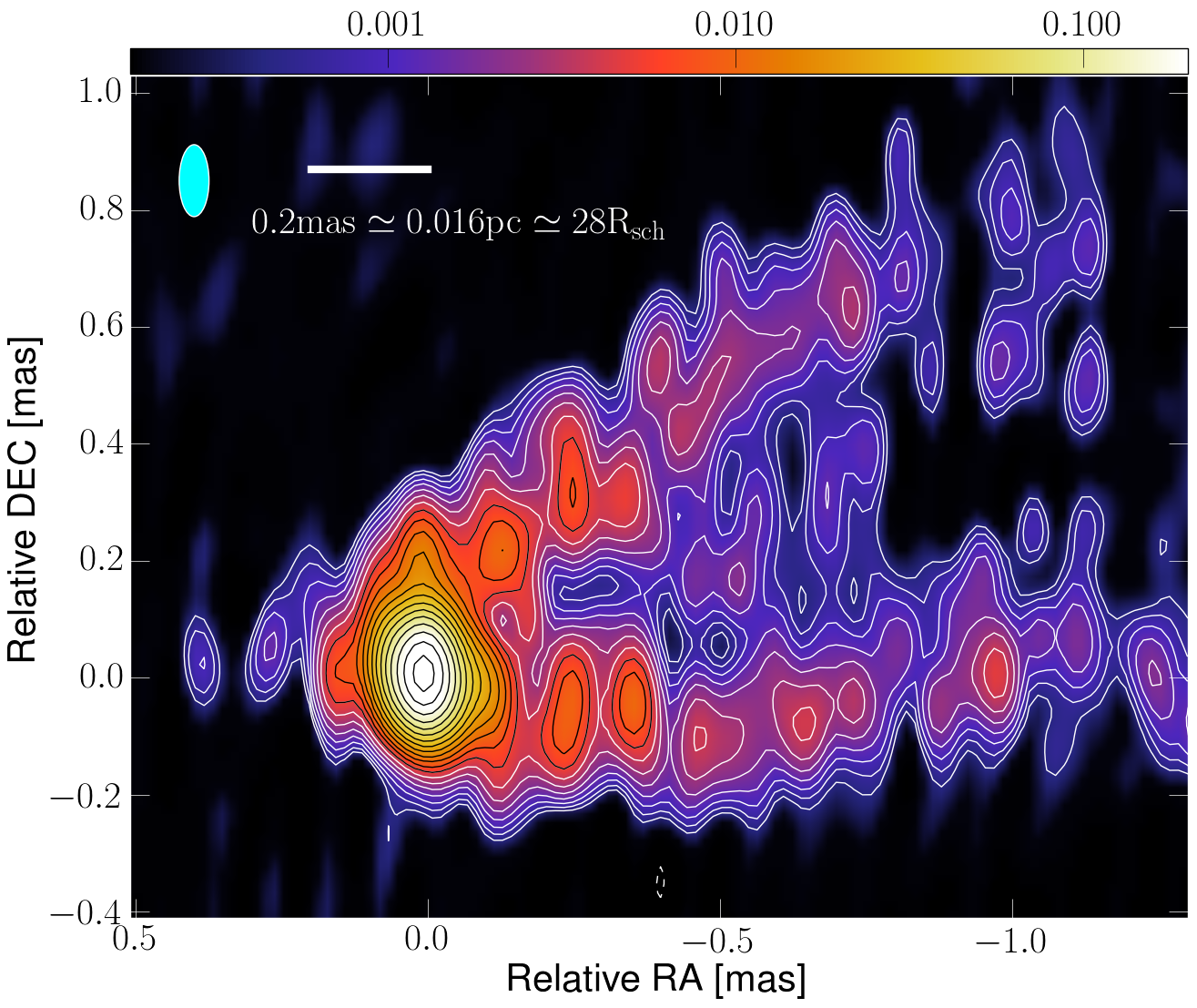}
 \caption{VLBI observations of M\,87 at 86\,GHz. The image was created after stacking 5 VLBI images taken during 2004-2015. (Image reproduced with permission from \cite{2016Galax...4...39K}, copyright by the authors)}
\end{figure}

The acceleration and collimation region could also be imaged in Cygnus\,A at 43\,GHz and 86\,GHz \citep{2016A&A...585A..33B, 2016A&A...588L...9B}. A multi-scale representation of this prototypical source is shown in Fig. 11. The observed kinematic properties in this powerful radio galaxy are quite similar to those inferred for M\,87: the flow is transversely stratified and mildly relativistic ($1<\Gamma<2.5$), it has a parabolic shape up to scales of ${\sim}10^3-10^4$\,$R_S$, and shows acceleration with two distinct regimes, a fast regime (with Lorentz factor increasing as fast as the jet radius) up to ${\sim}5000$\,$R_S$ and a slow regime up to
${\sim}10^4$\,$R_S$. At 86\,GHz, the two-sided jet could be resolved transversely also close to the apex, revealing a very wide jet base with a minimum width of $(227\pm98)$ $R_{\rm S}$. This value is much larger than the radius of the innermost stable circular orbit (ISCO) in the accretion disk ($1<R_{\rm ISCO}<9$ $R_{\rm S}$), implying that at least part of the flow must be anchored at large disk radii in the accretion disk. 

In this respect, Cygnus\,A differs from M\,87. Pilot observations of M\,87 performed at 1\,mm with the Event Horizon Telescope (EHT) \citep{2012Sci...338..355D} could determine an upper limit for the transverse size of the jet apex of only 5.5 $R_{\rm S}$, compatible with later findings by \cite{2014evn..confE..13K}. Unlike in Cygnus\,A, the initial jet width in M\,87 is comparable with the radius of the innermost stable circular orbit (ISCO), suggesting that the jet is launched from the inner regions of the disk or from the ergosphere. Very similar conclusions were reached by \cite{2016A&A...595A..54M} through the analysis of the jet rotation at 43\,GHz.

\begin{figure}[!h]
\centering
 \includegraphics[width=0.8\textwidth]{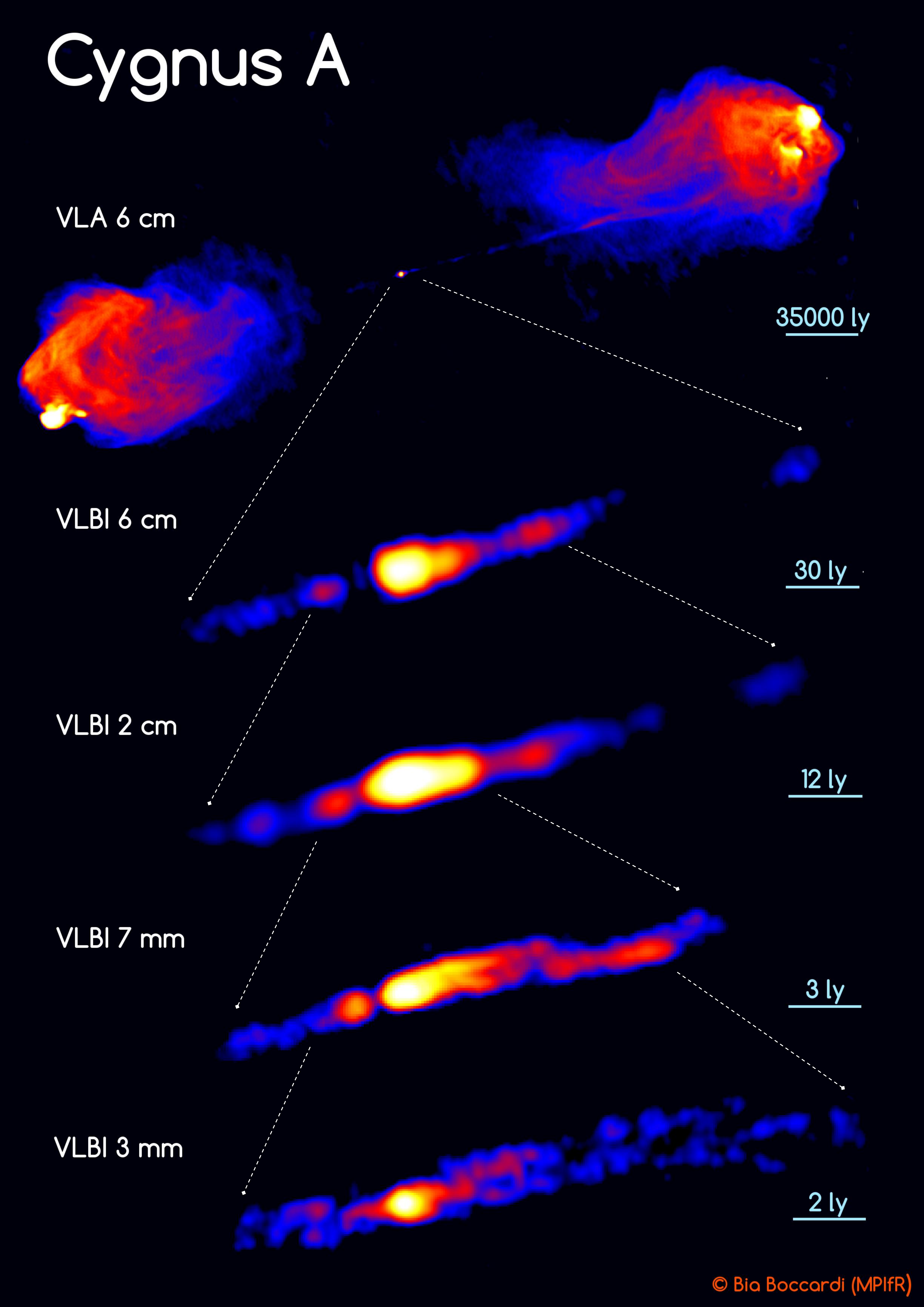}
 \caption{The radio galaxy Cygnus\,A on scales from hundreds of kilo-parsecs imaged with the Very Large Array \citep{1984ApJ...285L..35P} to the sub-parsec probed with mm-VLBI \citep{2016A&A...585A..33B, 2016A&A...588L...9B}. The VLBI images are created after stacking several epochs. Data at 2 cm are from the \textit{MOJAVE} survey \citep{2009AJ....137.3718L}. }
\end{figure}

All of the aforementioned results agree well with the theoretical predictions for magnetically dominated jets, supporting the currently most favored paradigm. 
Additional observational grounds favoring the magnetic launching scenario come from estimates of the magnetic field strength at the jet apex. 
In a recent analysis of a large number of sources from the \textit{MOJAVE} survey \citep{2014Natur.510..126Z}, the nuclear magnetic field strength has been derived from the jet radio luminosity. The inferred forces are extremely high, and comparable with the black hole's gravitational pull. Magnetic field strengths of the order of ${\sim}10^3$\,$\rm G$ or more were also derived in GMVA studies of OJ\,287 \citep{2017A&A...597A..80H} and NGC\,1052 \citep{2016A&A...593A..47B}, and may be compatible with the already discussed ALMA results presented by \cite{2015Sci...348..311M} for PKS 1330$-$211.
The existence of the so-called \textit{Magnetically Arrested Disks} (MADs), i.e., disks featuring dynamically important magnetic fields, was theorized by \cite{2003PASJ...55L..69N}. The adjective ``arrested'' refers to the accretion, which can be suppressed if the magnetic pressure gets strong enough, in the innermost regions of the disk. 

Numerical simulations \citep{2011MNRAS.418L..79T} confirm that MAD systems can produce strong jets, with powers
up to 140\% of the accreted rest-mass energy in the case of a maximally spinning black hole.  Simply speaking, this means that ``more than what comes in goes out'', and this surplus must be at the expenses of the black hole's rotational energy. This theoretical prediction agrees very well with another recent observational study of a large sample of blazars \citep{2014Natur.515..376G}, showing that the jet luminosity is correlated with the disk luminosity but  at the same time is much larger than it. Again, the contribution of a spinning black hole is necessary to account for the jet power. These results indicate that the Blandford-Znajeck process is appropriate for reproducing the properties of relativistic jets in blazars. If, on the other hand, the Blandford-Payne mechanism is also at play, but produces less powerful jets, it is natural to expect that the black-hole driven component will be dominant due to the stronger boosting. The observation of disk winds in highly misaligned sources like Cygnus\,A is 
well placed in such a scenario. 

A deep investigation of other objects in Table 2 is ongoing. In the future, the possibility of enlarging the sample of sources where the jet formation region can be imaged will depend on the compactness of the jet bases at frequencies higher than 86\,GHz. Results from the surveys \citep{2000A&A...364..391L, 2008AJ....136..159L, 2016cosp...41E.710G} indicate a decrease of brightness temperature with increasing frequency, which may hinder the detection of the innermost jet emission at 230 GHz and higher. On the other hand, experiments planned for the near future will also profit from the deployment of the phased ALMA and from the improved sensitivity of global arrays, which may enable to probe the jet formation region in fainter objects.

\subsection{\large Sagittarius A*}

Sagittarius A* (Sgr A*) is one of the three main components forming the Sagittarius A complex, located at the center of the Milky way.
This compact object, emitting from radio to $\gamma$-rays with a peak at sub-millimeter wavelengths, is believed to mark the position of the supermassive black hole at the center of our galaxy. With a mass of about 4 millions solar masses \citep{2009ApJ...692.1075G} and a distance of $\sim$8.3 kiloparsecs \citep{2014ApJ...783..130R}, Sgr A* is characterized by the largest apparent event horizon size of any black hole candidate in the Universe, which makes it a privileged cosmic laboratory for studies aimed at resolving the innermost accretion flow on Schwarzschild radius scales. The latter is not only important for understanding the accretion mechanisms and the formation of possible outflows, but also for verifying fundamental physics. On such scales, strong gravity effects are expected to be at play, and crucial tests of general relativity may be performed \citep{2016PhRvL.116c1101J, 2016PhRvL.117i1101J}. 

In the radio band, the galactic center is characterized by a nearly flat radio spectrum with a flux density of approximately 1 Jansky. The source is compact and well detected on VLBI scales. Since the first single-baseline detection at 215\,GHz in 1995 \citep{1997A&A...323L..17K, 1998A&A...335L.106K}, several test observations of Sgr A* have been conducted in this band. While successful VLBI experiments have been performed also at 43 and 86\,GHz \citep[e.g.,][]{1993A&A...274L..37K, 2011A&A...525A..76L, 2016A&A...587A..37R, 2016ApJ...824...40O}, observing at even higher frequencies is essential in this object in order to partially overcome the strong broadening caused from interstellar scattering. This affects quite heavily the emission at 43\,GHz and 86\,GHz, and is still of the order of $20$ $\rm \mu$as at 215\,GHz \citep{2006ApJ...648L.127B}. 
In a three-station VLBI experiment at the latter frequency, a scattering-corrected source size of 37 $\mu$as could be determined \citep{2008Natur.455...78D}. Assuming the aforementioned distance and mass parameters, the radius of the black hole event horizon subtends an angular scales of ${\sim}10$ $\rm \mu as$, but the apparent horizon is expected to be enlarged due to gravitational lensing, to a diameter of ${\sim}52$ $\rm \mu as$. Therefore, quite surprisingly, the size measured at 215\,GHz was found to be smaller than the apparent dimension of the event horizon. This remarkable result suggests that the emission from Sgr\,A* may not be centered on the exact location of the black hole but it may originate in its surroundings. 

Several models have been elaborated with the aim of identifying the nature of the emission region. Many invoke the emission from an inefficient accretion flow, such as the ADAF \citep[advection-dominated accretion flow,][]{1994ApJ...428L..13N} and the ADIOS \citep[advection-dominated inflow-outflow solutions,][]{1999MNRAS.303L...1B}. While these models may explain the observed emission up to the high energies, they have been found to significantly under-predict the low-frequency radio continuum \citep[e.g.,][]{2002A&A...383..854Y}. The latter component can be better reproduced by including the contribution of hot electrons ($T\sim10^{11}$ $\rm K$), often proposed to originate in a weak relativistic jet coupled to the accretion flow 
\citep[e.g.,][]{2000A&A...362..113F}. With a bolometric luminosity several orders of magnitude below the Eddington limit (${\sim}10^{-8}L_{\rm Edd}$), Sgr A* is highly under-luminous when compared to a classical AGN. The study of this putative jet is therefore fundamental for exploring the jet formation phenomenon under different physical conditions. 

In the near future, the Event Horizon Telescope will be capable of producing detailed imaging of the radio structure in Sgr A*, possibly shedding light on the nature of the emission. In the meantime, the visibility and closure phase data collected so far at 230 GHz are already providing important constraints.
The radio emission has been observed to vary on time scales of days \citep{2011ApJ...727L..36F}, and a persistent asymmetry in the source structure has been deduced from the non-zero closure phases \citep{2016ApJ...820...90F}. Moreover, a recent study has shown that the radio emission is highly polarized, up to ${\sim}70\%$, on scales of 6\,$\rm R_S$ \citep[][Fig. 12]{2015Sci...350.1242J}, and that the polarization direction varies spatially. The polarization degree decreases by one order of magnitude when looking at the short baselines, to values that match very well those measured at the same frequency on arcsecond scales \citep[e.g.,][]{2003ApJ...588..331B}. This comparison shows, on the one hand, that the polarization arises in the very central regions of the source, and is not associated to dust emission in the surroundings. Secondly, it points out the absolute importance of resolving these compact regions for unveiling the coherent polarization structures in the source. There is no single explanation 
for the existence of strong ordered magnetic fields in Sgr A*. They could signal the presence of a magnetically dominated region formed after accumulation of magnetic flux near the event horizon, similarly to that inferred for the cores of extragalactic jets (Sect. 4.4), or they could arise as a consequence of instabilities 
\citep[e.g., the magnetorotational instability - MRI, ][]{1991ApJ...376..214B} in the magnetized and rotating accretion disk. Finally, they could also be associated with the emergence of a relativistic jet.  

\begin{figure}
\centering
 \includegraphics[width=0.9\textwidth]{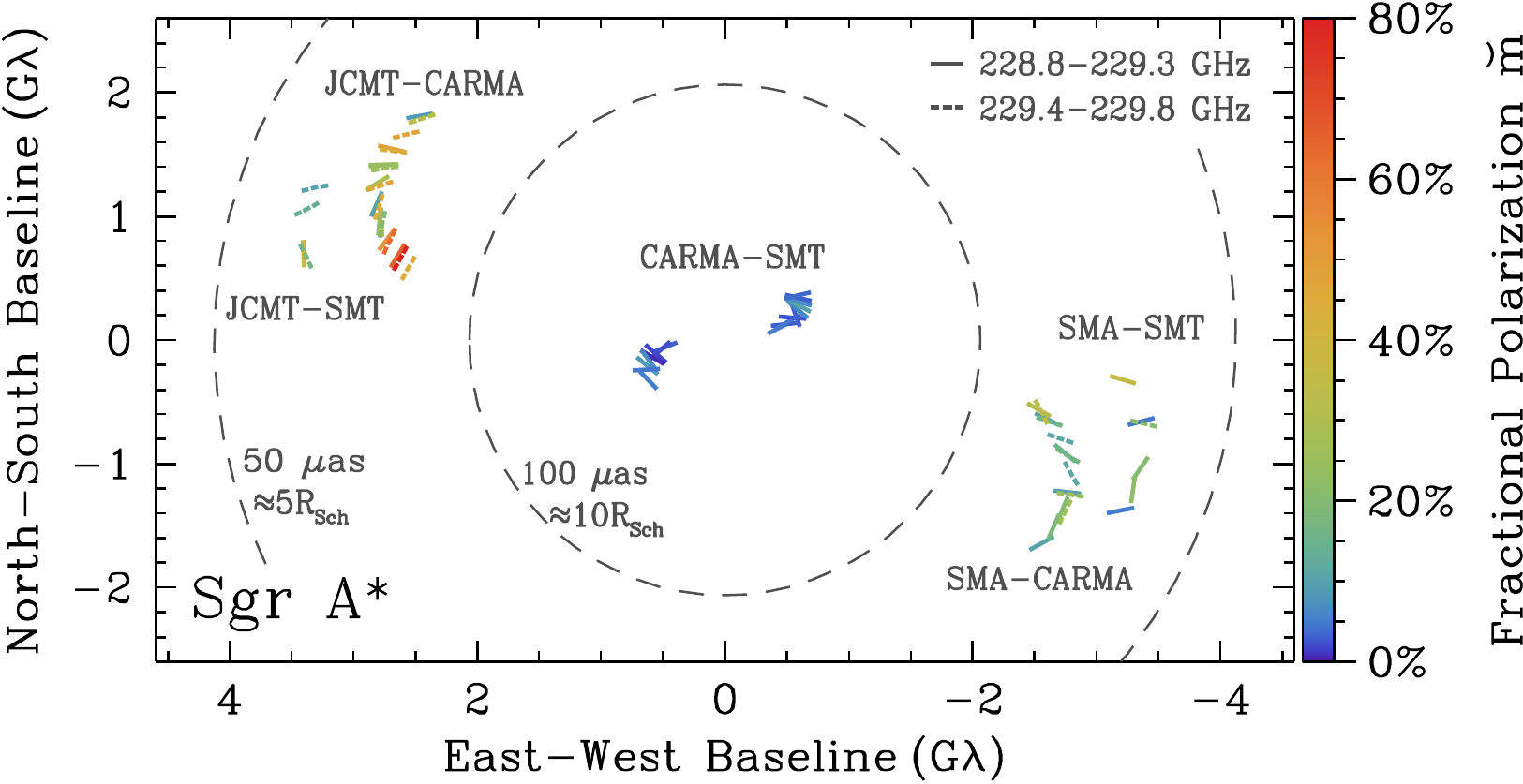}
 \caption{Interferometric fractional polarization measurements for Sgr A* during EHT observations in 2013. Polarization degrees up to $70\%-80\%$ are observed. The telescopes participating to the experiment were the phased CARMA (Combined Array for Research in Millimeter-wave Astronomy) in California, the SMT (Submillimeter Telescope) in Arizona, the SMA (Submillimeter Array) and the JCMT (James Clerk Maxwell Telescope) in Hawaii. (Image reproduced with permission from \cite{2015Sci...350.1242J}, copyright by AAAS)}
\end{figure}

The second and most crucial objective of the EHT project will be the direct imaging of the so-called ``black hole shadow''. The shadow is the dark region around a black hole \citep{1973blho.conf..215B, 1979A&A....75..228L, 2000ApJ...528L..13F}, expected to be observable in the presence of a background source, e.g., an optically thin plasma. The shadow should be surrounded by a thin photon ring, whose shape and size are independent on the physical detail of the accretion process and should only be determined by the basic properties of the black hole, i.e., its mass and spin, and by its inclination. The photon ring is predicted to be perfectly circular in the case of a non-spinning black hole with spherical symmetry, while crescent-like morphologies should arise in the case of a Kerr black hole.
Concerning the brightness distribution around the photon ring, this is expected to be asymmetric in all cases due to differential Doppler beaming, but its detailed characteristics will also depend on the physical conditions of the accretion flow \citep[see e.g.,][]{2009ApJ...706..497M, 2010ApJ...717.1092D}. 
Ultimately, the analysis of the geometrical and physical properties of the shadow will provide further constraints on the properties of the plasma flow in SgrA* and, most importantly, a direct test for the validity of the no-hair theorem and of the Einstein's theory of gravity in general.
 
Through a recent four-site experiment at 230 GHz, providing a resolution of ${\sim}2.5$ Schwarzschild radii, \textcolor{blue}{Lu et al.} (\textcolor{blue}{in prep.}) were able to resolve the event horizon scale structure of Sgr\,A*, which revealed a complex brightness distribution compatible with the expectations for a crescent-like structure. VLBI observations in the near future should allow more detailed imaging. Upcoming experiments will offer new challenges to observers, due to the sparse $(u,v)$-coverage, influence of scattering in the images, and time variability during the observations. Several methods are being developed, concerning non-imaging analysis and addressing the mitigation of those effects \citep{2014ApJ...795..134F, 2015ApJ...805..180J, 2016ApJ...833...74J, 2016ApJ...817..173L}.

The investigation of the Galactic center is considered the main scientific goal of the Event Horizon Telescope (EHT) project, whose efforts are shared with --and complemented by-- the \textit{BlackHoleCam} European collaboration \cite[e.g.,][]{2017IJMPD..2630001G}. In this context, vibrant research activity is being carried out at present on multiple fronts, from the refinement of theoretical models and simulations to the enhancement of VLBI arrays and data analysis techniques. For a recent overview of the EHT efforts concerning these subjects, the reader is referred to \cite{2016Galax...4...54F}.

\section{\large The future of mm-VLBI}
Fifty years after its birth, VLBI is experiencing a second youth due to its development in the millimeter and sub-millimeter regimes. Unexplored frontiers have now become accessible thanks to it, new science has been produced and more fundamental results are foreseen in the coming years. 

The GMVA is conducting successful biannual observations at 86\,GHz, and has provided important constraints for theories describing blazar physics and jet formation. In the future, a more reliable description of the jet kinematic properties and variability could be obtained through a high-cadence monitoring which, given the short timescales characterizing blazars, is currently too sparse.
Significant improvements could also come by implementing multi-frequency receiving systems, following the example given by the Korean VLBI Network \citep{2017arXiv170504776A}. Besides the obvious advantages it would provide for the accurate characterization of the jet spectral properties, this method also enables to enhance the data quality. By applying a technique known as \textit{source-frequency phase referencing}, the measurements obtained at the lowest frequencies, which are less affected by atmospheric opacity, can in fact be used to correct the phase at higher frequencies. 

In addition to the Korean VLBI Network, which is at present regularly participating in the observing sessions, other telescopes may enhance the future imaging capabilities at 86\,GHz. These include the NOEMA (NOrthern Extended Millimeter Array), successor of the Plateau du Bure observatory, the Large Millimeter Telescope in Mexico, the phased ALMA (at present commissioned for observing at 86\,GHz and 230 GHz, potentially capable of going down to 43\,GHz and up to 345 GHz and even beyond), and the Greenland Telescope, starting its commissioning in early 2017. These latter sites are also part of the Event Horizon Telescope, together with APEX in Chile, the JCMT (James Clerk Maxwell Telescope) and the SMA (Sub-Millimeter Array), both located in Hawaii, the Arizona Radio Observatory (SMT), the SPT (South Pole Telescope), and the 30-m telescope atop Pico Veleta in Spain.

The ALMA Phasing Project \citep[e.g.,][]{2014AAS...22344302M}, led by the Haystack Observatory of the Massachusetts Institute of Technology in collaboration with several institutions worldwide, started in 2011 and completed commissioning in 2015.  Its goal was to phase-up the 64 ALMA dishes to form a single VLBI station with equivalent diameter of 84 meters. Additionally to the phasing, the project includes polarization conversion to match the typical circular polarization recorded by the VLBI stations with the X/Y (linear) polarization used by ALMA \citep{2016A&A...587A.143M}.  

ALMA has performed its first regular VLBI observations in April 2017 during Cycle 4. The GMVA was assigned the role of the network provider at 86\,GHz, and the Event Horizon Telescope consortium (supervised by the National Radio Astronomy Observatory) this role at 230\,GHz. Thanks to its large collecting surface, ALMA is expected to provide a substantial boost in sensitivity, since the root-mean-square noise of the image will be typically reduced by a factor of three. The first observations including ALMA are still limited by technical reasons, such as a flux density threshold of 500\,mJy and the need of long integration time for polarization calibration. Those limitations will be loosen in the future, for instance, when a sky model can be introduced to the phasing. 

The possibility of achieving an excellent sensitivity will be crucial in future EHT experiments, aiming at producing a direct image of the black-hole shadow. High sensitivity and good image fidelity will be obtained not only through the expansion of the array, which after 2020 should also include a telescope in Namibia, Africa (AMT - Africa Millimeter Telescope)\footnote{http://www.ru.nl/rbhh-eng/}, but also thanks to the improvement of the recording capabilities at each site, with subsequent increase in bandwidth. Recording capabilities are now boosted by the use of digital techniques and, from the 512 $\rm{Mbit/s}$ used several years ago, present developments in digital base band conversion \citep[e.g.,][]{2012SPIE.8452E..2WT} bring them to 64\,$\rm{Gbit/s}$, yielding instantaneous bandwidths of 16\,GHz. 

Millimeter VLBI is an example of cutting edge technology at the service of science. The international effort recently spent in its further upgrade will enable the exciting scientific questions presented in this article to be fully addressed, by zooming into some of the most violent phenomena in the Universe. Landmark tests for our current knowledge of fundamental physics will be provided, together with final and direct evidence for the existence of black holes at the center of galaxies.

 \begin{acknowledgements}
Open access funding provided by Max Planck Society. The authors would like to thank the anonymous referee and the editors for their suggestions, which significantly improved the article. The authors are also thankful to Andrei Lobanov, Rebecca Azulay, Carolina Casadio, Vassilis Karamanavis,
Ioannis  Myserlis,  Alice  Pasetto,  and  Thalia  Traianou  for  reading  and  commenting  on  the  manuscript, and  to  Uwe  Bach  for  kindly  providing  the  calibrated  6  cm  data  of  Cygnus  A  shown  in  Fig.  11.  E.R. was partially supported by the Spanish MINECO grant AYA2015-63939-C2-2-P and by the Generalitat Valenciana grant PROMETEOII/2014/057. This study makes use of 43 GHz VLBA data from the VLBA-BU Blazar Monitoring Program (VLBA-BU-BLAZAR; http://www.bu.edu/blazars/VLBAproject.html), funded by NASA through the Fermi Guest Investigator Program. The VLBA is an instrument of the Long Baseline Observatory. The Long Baseline Observatory is a facility of the National Science Foundation operated by Associated Universities, Inc. 
This research has made use of data from the \textit{MOJAVE} database that is maintained by the \textit{MOJAVE} team \citep{2009AJ....137.3718L}.
The European VLBI Network is a joint facility of European, Chinese, South African, and other radio astronomy institutes funded by their national research
councils. 
 \end{acknowledgements}

\def\memsai{Memorie della Società Astronomica Italiana}
\def\aj{AJ}%
\def\actaa{Acta Astron.}%
\def\araa{ARA\&A}%
\def\apj{ApJ}%
\def\apjl{ApJ}%
\def\apjs{ApJS}%
\def\ao{Appl.~Opt.}%
\def\apss{Ap\&SS}%
\def\aap{A\&A}%
\def\aapr{A\&A~Rev.}%
\def\aaps{A\&AS}%
\def\azh{AZh}%
\def\baas{BAAS}%
\def\bac{Bull. astr. Inst. Czechosl.}%
\def\caa{Chinese Astron. Astrophys.}%
\def\cjaa{Chinese J. Astron. Astrophys.}%
\def\icarus{Icarus}%
\def\jcap{J. Cosmology Astropart. Phys.}%
\def\jrasc{JRASC}%
\def\mnras{MNRAS}%
\def\memras{MmRAS}%
\def\na{New A}%
\def\nar{New A Rev.}%
\def\pasa{PASA}%
\def\pra{Phys.~Rev.~A}%
\def\prb{Phys.~Rev.~B}%
\def\prc{Phys.~Rev.~C}%
\def\prd{Phys.~Rev.~D}%
\def\pre{Phys.~Rev.~E}%
\def\prl{Phys.~Rev.~Lett.}%
\def\pasp{PASP}%
\def\pasj{PASJ}%
\def\qjras{QJRAS}%
\def\rmxaa{Rev. Mexicana Astron. Astrofis.}%
\def\skytel{S\&T}%
\def\solphys{Sol.~Phys.}%
\def\sovast{Soviet~Ast.}%
\def\ssr{Space~Sci.~Rev.}%
\def\zap{ZAp}%
\def\nat{Nature}%
\def\iaucirc{IAU~Circ.}%
\def\aplett{Astrophys.~Lett.}%
\def\apspr{Astrophys.~Space~Phys.~Res.}%
\def\bain{Bull.~Astron.~Inst.~Netherlands}%
\def\fcp{Fund.~Cosmic~Phys.}%
\def\gca{Geochim.~Cosmochim.~Acta}%
\def\grl{Geophys.~Res.~Lett.}%
\def\jcp{J.~Chem.~Phys.}%
\def\jgr{J.~Geophys.~Res.}%
\def\jqsrt{J.~Quant.~Spec.~Radiat.~Transf.}%
\def\memsai{Mem.~Soc.~Astron.~Italiana}%
\def\nphysa{Nucl.~Phys.~A}%
\def\physrep{Phys.~Rep.}%
\def\physscr{Phys.~Scr}%
\def\planss{Planet.~Space~Sci.}%
\def\procspie{Proc.~SPIE}%

\scriptsize
\bibliographystyle{apalike}        
\bibliography{references.bib}   

\end{document}